\documentclass[njp,10pt]{iopart}
\usepackage{graphicx}
\expandafter\let\csname equation*\endcsname\relax
\expandafter\let\csname endequation*\endcsname\relax
\usepackage{graphicx}
\usepackage{amsmath}
\usepackage{amsfonts}
\usepackage{amssymb}
\usepackage{amsthm}
\usepackage{amstext}
\usepackage{soul}
\usepackage{dsfont}
\usepackage[utf8]{inputenc}
\usepackage[hidelinks]{hyperref}
\usepackage[caption=false]{subfig}
\usepackage{epstopdf}
\usepackage{cellspace}
\hypersetup{colorlinks=true, citecolor=blue, urlcolor=blue, linkcolor=blue}
\usepackage{tabularx,ragged2e,booktabs}
\usepackage[dvipsnames]{xcolor}
\usepackage{braket}
\usepackage{ulem}
\usepackage{color}
\usepackage[numbers,sort&compress]{natbib}

\usepackage[left=1in,right=1in,top=1in,bottom=1in,footskip=.25in]{geometry}

\renewcommand\vec{\boldsymbol}

\def\f12{\frac{1}{2}}
\newcommand{\be}{\begin{equation}}
\newcommand{\ee}[1]{\label{#1} \end{equation}}

\usepackage{dcolumn}
\usepackage{bm}
\usepackage{cellspace}

\usepackage{etoolbox}
\makeatletter
\def\@mkboth#1#2{}
\newlength\appendixwidth
\preto\appendix{\addtocontents{toc}{\protect\patchl@section}}
\newcommand{\patchl@section}{%
  \settowidth{\appendixwidth}{\textbf{Appendix }}%
  \addtolength{\appendixwidth}{1.5em}%
  \patchcmd{\l@section}{1.5em}{\appendixwidth}{}{\ddt}%
}
\makeatother

\appto\appendix{\addtocontents{toc}{\protect\setcounter{tocdepth}{1}}}

\appto\listoffigures{\addtocontents{lof}{\protect\setcounter{tocdepth}{1}}}
\appto\listoftables{\addtocontents{lot}{\protect\setcounter{tocdepth}{1}}}

\usepackage{mnsymbol}

\newcommand{\RN}[1]{\textup{\uppercase\expandafter{\romannumeral#1}}}

\newcommand{\kket}[1]{\left|#1\right\rrangle}

\renewcommand{\emph}[1]{\textit{#1}}

\begin{document}
\title{Solving quantum dynamics with a Lie algebra decoupling method}
\author{Sofia Qvarfort$^{1,2}$ and Igor Pikovski$^{2,3}$}

\address{$^1$ Nordita, KTH Royal Institute of Technology and Stockholm University, Hannes Alfv\'{e}ns v\"{a}g 12, SE-106 91 Stockholm, Sweden \\
$^2$ Department of Physics, Stockholm University, AlbaNova University Center, SE-106 91 Stockholm, Sweden \\
$^3$ Department of Physics, Stevens Institute of Technology, Castle Point on the Hudson, Hoboken, New Jersey 07030, USA }
\ead{sofia.qvarfort@fysik.su.se, igor.pikovski@fysik.su.se}

\date{\today}

\begin{abstract}
At the heart of quantum technology development is the control of quantum systems at the level of individual quanta. Mathematically, this is realised through the study of Hamiltonians and the use of methods to solve the dynamics of quantum systems in various regimes. Here, we present a pedagogical introduction to solving the dynamics of quantum systems by the use of a Lie algebra decoupling theorem. As background, we include an overview of Lie groups and Lie algebras aimed at a general physicist audience. We then prove the theorem and apply it to three well-known examples of linear and quadratic Hamiltonian that frequently appear in quantum optics and related fields. The result is a set of differential equations that describe the most Gaussian dynamics for all linear and quadratic single-mode Hamiltonian with generic time-dependent interaction terms. We also discuss the use of the decoupling theorem beyond quadratic Hamiltonians and for solving open-system dynamics.
\end{abstract}

\maketitle
\tableofcontents


\section{Introduction}
Quantum physics results in unique phenomena, such as quantum superpositions and entanglement, that can be harnessed for applications ranging from quantum technologies to tests of fundamental physics. However, developing such applications requires exact knowledge and control of individual quanta. As new experimental platforms and hybrid systems emerge, novel dynamics and interaction regimes become increasingly accessible.
In order to fully unlock the potential of these systems, it is crucial to be able to model their quantum dynamics. 
To mention just a few examples, quantum dynamics plays a crucial role for quantum control~\cite{dong2010quantum}, quantum information processing~\cite{blais2007quantum}, and quantum sensing~\cite{degen2017quantum}.

Beyond quantum technologies, searches for new effects in fundamental physics often result in prediction for changes in quantum dynamics. To detect these often extremely weak effects, it is crucial to be able to model the system dynamics exactly. 
As an example, the search for a quantum theory of gravity has resulted in the study of modifications to the usual algebras used in quantum theory~\cite{maggiore1993generalized, kempf1995hilbert}. Such a modified algebra necessarily leads to changes in the dynamics of quantum systems. In order to predict observable effects, we typically require methods to construct the resulting decoupled unitary operations, at least perturbatively~\cite{pikovski2012probing}.

It can however be challenging to treat quantum dynamics analytically. The core of the difficulty  lies in the non-commutativity of operators that enter into the Hamiltonian. While it is always possible to write down the formal solution to the dynamics in terms of an exponential operator, this operator cannot always be tractably applied to the initial quantum state. The notion of solving quantum dynamics can therefore be understood as finding a closed-form expression of the time-evolution operator that facilitates the computation of any quantity of the system. Ultimately, this is equivalent to solving Schrödinger’s differential equation. 

In many fields of quantum science, it is often sufficient to use approximations such as considering only small quantum perturbations around classical solutions, or to average over many systems. As long as the effects of interest are weak, such perturbative solutions can be used to accurately model the behaviour of experiments in the laboratory. 
Many mathematical methods have been developed to express, manipulate and truncate exponential operators~\cite{zassenhaus1939verfahren, magnus1954exponential, wei1964global, kumar1965expanding, wilcox1967exponential, hatano2005finding}. In order to control the full quantum behaviour of individual quantum systems, however, such as in quantum optics, quantum information and for quantum technologies, it can be necessary to go beyond these approximations.

One mathematical technique that can be used to solve quantum dynamics exactly makes use of the Lie algebra induced by the Hamiltonian. The method is based on a Lie algebra decoupling theorem, which was put forwards by Wei and Norman in 1963~\cite{wei1963lie}. At its core is the observation that a finite Lie algebra generated by a Hamiltonian can be used as a basis that allows for a set of scalar differential equations to be derived. Should the resulting differential equations have analytic solutions, the system dynamics can be solved exactly. 

In this work, we provide a pedagogical introduction to the Lie algebra decoupling theorem and its application. Specifically, we apply the theorem towards solving the dynamics of three Hamiltonians that frequently appear in quantum optics and adjacent fields. The Hamiltonians we consider have both linear and quadratic interaction terms, where each interaction term depends on time in an arbitrary fashion. As such, the theorem allows us to provide a treatment of the most general case of Gaussian dynamics. 

The work is structured as follows. Section~\ref{sec:lie:algebra} provides a  mathematical introduction to Lie groups and Lie algebras. Section~\ref{sec:decoupling} contains a proof of the decoupling theorem, as well as a straight-forward recipe for how it can be applied. Then, in Sections~\ref{sec:linear:terms} and~\ref{sec:quadratic:drive}, we apply the theorem to solve the dynamics of a Hamiltonian with time-dependent linear and quadratic interaction terms, respectively. There are also ways in which these solutions can be combined to represent the most general Gaussian dynamics, which we demonstrate in Section~\ref{sec:linear:quadratic:drive}. The work is concluded with a discussion of symmetries and extensions to open-system dynamics in Section~\ref{sec:discussion}, as well as some final remarks in Section~\ref{sec:conclusions}. In the appendices that follow, we provide an extension of the decoupling theorem to phase space, as well as detailed calculations for each section in the main text.

\section{Introduction to Lie algebras and Lie groups} \label{sec:lie:algebra}

The Lie algebra decoupling theorem can be understood and applied without deep knowledge of Lie algebras, but for the interested reader, we here provide a basic mathematical introduction to Lie groups, Lie algebras, and the connection between them. In each case, we start the discussion with a formal definition, then provide a few examples that commonly appear in physics.

\subsection{Lie groups}
To understand Lie algebras, it helps to first introduce the concept of a Lie group. We start by recounting the  formal definition of a group. \vspace{0.2cm}

\noindent \textbf{Definition (group)}
A group $(G,*)$ is a set $G$ with group elements $g$ and a binary operation $*$ such that $G\times G \rightarrow G$, which satisfies three conditions:
\begin{enumerate}
\item \textit{Associativity}: For any three elements $x,y,z \in G$, we have $(x*y) *z = x* (y*z)$.
\item \textit{Identity}: The group must include an identity element $\epsilon \in G$. It must hold that $\forall g \in G$,  multiplying $g$ by the identity element leaves $g$ invariant, such that $\epsilon * g = g * \epsilon = g$.
\item \textit{Inverse}: For each element in the group, there must be an inverse element. That is, for each element $g \in G$ there is some inverse element $\bar{g} \in G$ such that $g * \bar{g} = \bar{g} * g = \epsilon$.
\end{enumerate}
Groups arise in many separate context in physics.
Perhaps one of the simplest examples of a group is the set of discrete rotations that leave a square invariant. That is, the group elements correspond to rotation actions (acting on a square) and the group operation corresponds to the composition of actions. This group is known as $Z_4$ and contains rotations of zero, $90^\circ$, $180^\circ$, and  $270^\circ$ degrees. This set of rotations satisfies the group axioms, since any combination of the group elements satisfies the same symmetry. That is, we can rotate the square by first $90^\circ$, then $180^\circ$, and it remains invariant.  The associativity criterion is satisfied by a proper representation of the group elements, such as a matrix. Lastly, the identity is equivalent to not rotating the square, and the inverse can easily be constructed by rotating the square in the reverse direction by the same amount of degrees.

We now proceed to discuss Lie groups. They are \textit{continuous} groups and play an ubiquitous role in physics and mathematics. For example, in quantum mechanics, the set of unitary time-evolution operators form a Lie-group, as we will see below.
We proceed with a formal definition of a Lie group. \vspace{0.2cm}

\noindent \textbf{Definition (Lie group)}:
A Lie group is a set $G$ with two structures:
\begin{enumerate}
\item $G$ is a group with the structure discussed in the definition of a group shown above.
\item $G$ is a smooth and real manifold. Smoothness means that the group operation and inverse map are differentiable. The group is described by a set of real parameters that describe the group elements.
\end{enumerate}

Examples of Lie groups include, for example, the translation group, the special unitary group SU(n), the group of all invertible linear maps, and the special orthogonal group in three dimensions SO(3). For all of these groups, the inverse and the zero elements can be constructed. The real line has continuous elements, where the inverse can be constructed by subtraction, and the identity element is zero. The special unitary group SU($n$) describes $n\times n$ unitary matrices with determinant $1$. Furthermore, the inverse element can be obtained through complex conjugation, and the identity is the $n \times n $ identity matrix. Finally, the SO(3) group describes the rotation of vectors in three dimensions. The inverse element can be constructed through orthogonality, and matrix multiplication automatically satisfies associativity.
In fact, most continuous rotation groups are Lie groups. A list of Lie groups can be found in Ref~\cite{fulton2013representation}.

\subsection{Lie algebras}

We are now finally ready to properly define a Lie algebra. Lie algebras are the key object of interest in this work, and the link between Lie groups and Lie algebras underpins the Lie algebra decoupling theorem. Given a Lie group, it is always possible to construct a Lie algebra from the group, and sometimes there are advantages to studying the algebra rather than the group itself.
We begin with the basic definition of a Lie algebra~\cite{gutowski2007symmetry}. 
\vspace{0.2cm}

\noindent\textbf{Definition (Lie algebra)}: A Lie algebra is a vector space $\vec{g}$ over some field $F$\footnote{A field is a fundamental algebraic structure in the form of a set, where where addition, subtraction, multiplication, and division are defined.}, together with a binary operation $[\cdot, \cdot] : g \times g \rightarrow g$ (the Lie bracket) which must satisfy the following axioms:
\begin{enumerate}
\item Bilinearity, such that $[ax + by, z]  = a[x,z] + b[y,z]$ and $[z, ax + by] = a[z, x] + b[z, y]$ for all scalars $a, b$ in the field $F$ and all elements $x,y,z$ in $g$.
\item Alternativity, which means that the Lie bracket is zero for the same element: $[x,x] = 0$.
\item The Jacoby identity, which states that
\begin{equation}
\left[ x, ]y,x] \right] + \left[ z, [x,y] \right] + \left[ y, [z, x] \right] = 0
\end{equation}
\end{enumerate}
We find that the commutator bracket $[A, B] = AB - BA$, which is commonly used in quantum physics, satisfies these criteria. In fact, the commutator bracket is often used and is a measure of how non-commutative an algebra is.

\subsection{Link between Lie groups and Lie algebras}
The next question is how Lie groups connect with Lie algebras. While this can be discussed to great mathematical detail, we here provide an example that is hopefully more intuitive to the quantum physicist. To begin with, we consider a Lie group $L$ with elements $G(\alpha) \in L$, where $\alpha$ is some real parameter. To determine the action of the element near identity, we can slightly perturb $G(\alpha)$ for a small $\alpha_j$ to find  
\begin{equation}
G(\alpha) \approx  + i \delta \alpha_j X_j  ,
\end{equation}
where we have defined the generator $\hat X_j$. Then, performing this small perturbation many times in addition to the identity operation, we find 
\begin{equation} \label{eq:exponential:map}
 \sum_k^\infty  \left( 1 + 	\frac{i \alpha_j X_j}{k} \right)^k \equiv e^{i \alpha_j X_j} = D(\alpha) .
\end{equation}
We can now also define the generator $X_j$ as the rate of change with respect to the parameter $\alpha_j$:
\begin{equation}
X_j \equiv - i \frac{\partial}{\partial \alpha_j} D(\alpha) \biggl|_j, 
\end{equation}
We note that Eq.~\eqref{eq:exponential:map} is, in fact, the definition of the exponential map. The $X_j$ are generators, which form a Lie algebra. The Lie algebra then generates the group together with the real parameters $\alpha_j$. In other words, given a Lie algebra with a set of $n$ elements one can always use the exponential map to generate a Lie group.

\section{The Lie algebra decoupling theorem} \label{sec:decoupling}

Equipped with some knowledge of Lie groups and Lie algebras, we are now ready to study the Lie algebra decoupling theorem, first outlined by Wei and Norman in 1963~\cite{wei1963lie}. This section closely follows the proof first developed in Ref~\cite{wei1963lie}, but with slightly different notation in order to be consistent with modern conventions in quantum information and quantum optics. See also Ref~\cite{qvarfort2020quantum} for a presentation of these methods in the context of optomechanical systems. For convenience, we set $\hbar = 1$ in this section.

Intuitively, the Lie algebra decoupling methods separates a dynamical problem into the notion of \emph{directions} of evolution (where the directions are defined by the algebra), from the \emph{speed} of the evolution (the amount with which each algebra element is applied to the quantum state). For example, in continuous variables quantum information, rotation, displacement and squeezing operators are often used to describe the trajectory of a quantum state in phase space.

The Lie algebra decoupling method effectively transforms the problem of solving an operator-valued linear differential equation into that of solving a coupled system of differential equations of real coefficients. One  advantage of using this method is that problems which would have required the use of large numerical Hilbert spaces can instead be treated by solving a set of potentially coupled scalar differential equations. While these equations do not always have analytic solutions and might similarly have to be solved using numerical methods, errors due to the limited size of numerical Hilbert spaces can be avoided.

In short, the Lie algebra decoupling method is concerned with solving the Schrödinger equation. Consider the first-order differential equation
\begin{equation} \label{chap:decoupling:eq:to:solve}
\frac{d \hat U(t)}{dt} = - i  \hat H(t) \, \hat U(t) \, ,
\end{equation}
where $\hat H(t)$ is the Hamiltonian and $\hat U(t)$ is a time-evolution operator. The formal solution to $\hat U(t)$ is given by
\begin{equation} \label{eq:U:formal:solution}
\hat U(t) = \mathcal{T} \mathrm{exp}\left[ - \frac{i}{\hbar} \int^t_0 \mathrm{d}t' \, \hat H(t) \right],
\end{equation}
where $\mathcal{T}$ indicates time-ordering.

We then assume that the Hamiltonian $\hat H(t)$ can be written as a finite sum with $m$ terms\footnote{A Hamiltonian with an infinite number of unique terms would by extension also generate an infinite Lie algebra. The Lie algebra decoupling theorem holds finite Lie algebras, which is why we also assume that the initial Hamiltonian can be written as a sum over finite $m$.} 
of constant operators $\hat H_j$ and general time-dependent coefficients $G_j(t)$: 
\begin{equation}
\hat H(t) = \sum_{j = 1}^m G_j(t) \, \hat H_j  .
\end{equation}
The set $\{ \hat H_j \}$ with $j = 1, 2, \ldots , m$ reproduces the Hamiltonian $\hat H(t)$.  It can be extended to a larger set with $n \geq m$ elements by taking the commutator of the elements in $\{\hat H_j \}$ and adding the result to the set of Hamiltonian terms. One could then write the original Hamiltonian as:
\begin{equation}
\hat H(t) = \sum_{j = 1}^n G_j (t) \, \hat H_j ,
\end{equation}
where the coefficients with $j > m$ are set to zero. Noting that the sum can be extended in this way makes it easier to write down some relations further on.

The full set of Hamiltonian terms $\{\hat H_1 , \hat H_2, \hat H_3 , \cdots , \hat H_n\}$ then form a Lie algebra $L$ under commutation of dimension $n$.  In other words, we can find the full Lie algebra generated by $\hat H(t)$ by commuting all the elements $\hat H_i$ of $\hat H(t)$. The  Lie bracket is the commutator relation $[\hat H_i, \hat H_j] \equiv \hat H_i \hat H_j - \hat H_j \hat H_i$. The Lie algebra $L$ is constructed from all operators in $\hat H(t)$, plus all the Lie products
\begin{equation}
\left[ \hat H_{\alpha_1} , \left[ \hat H_{\alpha_2} , \left[ \hat H_{\alpha_3}, \cdots \left[ \hat H_{\alpha_{r-1}}, \hat H_{\alpha_r} \right] \cdots \right] \right] \right]  ,
\end{equation}
where $\alpha_i = 1 $ to $m$, plus all linear combinations of such products.

If, through consecutive commutation, we find a finite number of element, the Lie algebra is finite. This is always true if the $\hat H_j$ are finite-dimensional matrices. However, there are cases where the Lie algebra is infinite, for which commutation of two or more operators continuously produce new elements that are not already part of the algebra. In those cases, the dynamics can rarely be solved exactly.

We now show that the existence of such a finite Lie algebra $L$ enables the decoupling of the time-evolution operator $\hat U(t)$ into a product of $n$ operators, namely
\begin{equation} \label{chap:decoupling:eq:decoupled:U}
\hat U(t) = \hat U_1 (t) \, \hat U_2 (t) \, \cdots \hat U_n(t)  ,
\end{equation}
where each component operator $\hat U_j(t)$  is an operator satisfying
\begin{equation}
\frac{d}{dt} \hat U_j(t) = - i \dot{F}_j \, \hat H_j \, \hat U_j(t)  .
\end{equation}
and where the functions $F_j$ are the real functions that we wish to determine.

In quantum theory, the advantage of writing $\hat U(t)$ in the form in Eq.~\eqref{chap:decoupling:eq:decoupled:U} is that when the action of each $\hat U_j(t)$ is known, it becomes straight-forward to apply them to a quantum state in the Schrödinger picture, or an operator in the Heisenberg picture. The advantage of such a method over numerical solvers which use finite-dimensional matrices is significant, as the key task shifts from evolving the operator-valued $\hat U(t)$ to obtaining analytic expressions for scalar the $F_j$-functions.

We are now at a point where we can concisely state the Lie algebra coupling theorem. \vspace{0.2cm}

\noindent\textbf{Theorem (Lie algebra decoupling theorem)}
 Suppose that the linear operator $\hat H(t)$ can be expressed in the form
\begin{equation} \label{chap:decoupling:eq:hamiltonian:sum}
\hat H(t) = \sum_{j = 1}^m G_j(t) \hat H_j ,
\end{equation}
where $m$ is a finite integer and where the functions $G_j(t)$ are scalar functions of time $t$ and the $\hat H_j$ are time-independent  operators which live in a Hilbert space $\mathcal{H}$. 
The dimension of $\mathcal{H}$ can be either finite or infinite. Let the Lie algebra $L$ generated by $\hat H(t)$  be of finite dimension $n$. Then there exists a neighbourhood of $t = 0$ in which the solution of the equation
\begin{equation} \label{chapt2:eq:schrodinger:equation:definition}
\frac{d \hat U(t)}{dt} = - i \hat H(t) \, \hat U(t) \, ,
\end{equation}
with the initial condition $\hat U(0) = 1$ may be expressed in the form
\begin{equation} \label{chap:decoupling:time:evolution:ansatz}
\hat U(t) = \exp[ - i \, F_1(t) \, \hat H_1]\, \exp[ - i \, F_2(t) \, \hat H_2] \ldots \exp[ - i \, F_n (t) \, \hat H_n(t) ] \, ,
\end{equation}
where $\hat H_1, \hat H_2 , \cdots , \hat H_n$ is a basis for $L$ and the set $\{F_j(t)\}$ are scalar functions of time $t$. The functions $F_j(t)$ depend only on the Lie algebra $L$ and the initial functions $G_j(t)$.

The same decoupling of an evolution operator can also be performed for the symplectic matrices when considering the evolution under any quadratic Hamiltonian. We demonstrate this fact in~\ref{sec:phase:space}.

\subsection{Proof of the decoupling theorem}
Our goal is to prove  the Decoupling Theorem. It is based on two lemmas: The first is the well-known Baker-Campbell-Hausdorff Lemma and the second one concerns the closure of the Lie algebra. We begin with Lemma 1, which states: \vspace{0.2cm}

\noindent\textbf{Lemma 1. (Baker-Campbell-Hausdorff)}. If two operators $\hat X, \hat Y\in L$, then $e^{\hat X} \hat  Ye^{-\hat  X} \in L$ and
\begin{equation} \label{chap:decoupling:eq:BCH:result}
e^{\hat X} \hat  Y e^{-\hat X} = \hat Y + [\hat X,\hat Y] + \frac{1}{2!}[\hat X, [\hat X, \hat Y]]  + \frac{1}{3!} [\hat X, [\hat X, [\hat X, \hat Y]]] + \cdots  .
\end{equation}
We define the new operator $ad\hat X$, where $ad\hat X, \hat X \in L$ by the equation
\begin{equation}
(ad \hat X) \hat Y = [\hat X,\hat Y]  ,
\end{equation}
where $\hat Y \in L$. Then we define powers of this equation as the nested operators
\begin{equation}
(ad \hat X)^2 \hat Y = [\hat X, [\hat X, \hat Y]] ,
\end{equation}
and so on. Thus the Baker-Hausdorff formula can be stated as
\begin{equation}
e^{\hat X} \hat  Y e^{-\hat X} = (e^{ad \hat X})\hat Y  .
\end{equation}
\vspace{0.1cm}

\noindent\textbf{Proof of Lemma 1.}
We begin by defining a function
\begin{equation} \label{chap:decoupling:eq:BCH:series}
\hat F(a) = e^{a \hat X} \hat Y e^{- a \hat X} = \sum^\infty_{n = 0} \frac{1}{n!} \hat  C_n a^n   ,
\end{equation}
where the $\hat C_n$ are operator coefficients, which are independent of $a$. When $a = 1$, the coefficients correspond to the case we are considering.
Our goal is to derive expressions for these coefficients in the form of a recursion relation. We first note that
\begin{equation} \label{chap:decoupling:eq:BCH:differential}
\frac{d}{da} \hat F(a) = \left[ \hat X,  \hat F(a) \right] \, .
\end{equation}
Inserting Eq.~\eqref{chap:decoupling:eq:BCH:series}  into Eq.~\eqref{chap:decoupling:eq:BCH:differential} we find
\begin{equation}
\sum_{n = 1}^\infty \hat C_n \frac{1}{( n- 1)! } a^{n - 1} = \sum_{n = 0}^\infty \frac{1}{n!} \left[ X, \hat C_n \right] a^n   .
\end{equation}
The sum on the left-hand side can be rewritten by setting $n \rightarrow n + 1$, such that we find the formula
\begin{equation} \label{chap:decoupling:eq:BCH:coefficients}
\hat C_{n + 1} = \left[ \hat X, \hat C_n\right] a^n \, .
\end{equation}
This way, all coefficients can be generated through repeated commutation with $X$. We also have that $C_0 = Y$, which follows from simply Taylor expanding the exponentials in Eq.~\eqref{chap:decoupling:eq:BCH:series}. The coefficients in Eq.~\eqref{chap:decoupling:eq:BCH:coefficients} can then be used to generate all the coefficients in Eq.~\eqref{chap:decoupling:eq:BCH:result}.

The last lines in the lemma follow from the definition of $(ad X)$ as can be seen by writing
\begin{align}
e^{\hat X} \hat Y e^{- \hat X} &= \left( e^{ad \hat X} \right) \hat Y \nonumber \\
&= \sum_n \frac{1}{n!} (ad \hat X)^n \hat Y \nonumber \\
&=\hat  Y + [\hat X, \hat Y]  + \frac{1}{2!} \left[ \hat X, [\hat X, \hat Y] \right]  + \frac{1}{3!} \left[ \hat X, \left[ \hat X, [\hat X, \hat Y] \right] \right] + \ldots \, .
\end{align}
This concludes the proof of Lemma~1.
We proceed with the second lemma. \vspace{0.2cm}

\noindent\textbf{Lemma 2. (Lie algebra basis)}
Let $\hat H_1, \hat H_2 , \cdots , \hat H_n$ be a basis for the Lie algebra $L$. Then it follows that
\begin{align} \label{chap:decoupling:lemma:Lie:algebra:basis}
\left( \prod_{j = 1}^r \exp[ - i \, F_j \hat H_j] \right) \, \hat H_k \, \left( \prod_{j = r}^1 \exp[ i \, F_j \hat H_j ] \right) = - i \,  \sum_{j = 1}^n \xi_{jk} \hat H_j \, ,
\end{align}
where $r = 1, \cdots , n, $ and where each $\xi_{jk} \equiv \xi_{jk} (G_1, \cdots, G_r)$, is a function of all its arguments. \qedsymbol \vspace{0.2cm}

\noindent\textbf{Proof of Lemma~2.} Our goal is to establish that the $\xi_{jk}$ are analytic functions. This follows from repeatedly applying Lemma~1 to Eq.~\eqref{chap:decoupling:lemma:Lie:algebra:basis}. We demonstrate the first few lines of this proof.

Consider $r = 1$, which is the simplest case. We find, by using Eq.~\eqref{chap:decoupling:eq:BCH:result},
\begin{align} \label{chap:decoupling:eq:proof:lemma:2}
\exp[- i \,  F_1 \hat H_1 ] \hat H_k \exp[ i \, F_1 \hat H_1 ] =& \,  \hat H_k   - i \, F_1 [\hat H_1, \hat H_k] + \frac{( - i \, F_1)^2}{2!} [\hat H_1,[\hat H_1,  \hat H_k ]] \nonumber \\
&+ \frac{( - i \, F_1)^3}{3!} [\hat H_1, [\hat H_1, [\hat H_1, \hat H_k ]]] + \ldots  \, .
\end{align}
Now, since the Lie algebra is closed under commutation, it means that one of the terms eventually reads
\begin{equation}
[\hat H_1,  (\mathrm{ad} \hat H_1)^n \, \hat H_k] = [\hat H_1, \hat H_1]  = 0 ,
\end{equation}
for some integer $n$, which means that Eq.~\eqref{chap:decoupling:eq:proof:lemma:2} contains a finite number of terms with different powers of $F_1$.  Since all $F_1$ are analytic, the resulting functions $\xi_{jk}$ in the right-hand side of Eq.~\eqref{chap:decoupling:lemma:Lie:algebra:basis} are necessarily analytic. The same argument can be made for multiplication of additional terms when $r \neq 1$. This concludes the proof of Lemma~2. \qedsymbol

We are now ready to prove the Lie algebra decoupling theorem. The proof makes use of both Lemma~1 and 2. \vspace{0.2cm}

\noindent\textbf{Proof (Lie algebra decoupling theorem)}
We first note that we can write down an extended Hamiltonian
\begin{equation}
\tilde{\hat{H}}(t) = \sum_{j = 1}^n G_j(t) \, \hat H_i  \,\quad \quad \mbox{instead of} \quad \quad
\hat H(t) = \sum^m_{j = 1} G_j(t) \, \hat H_i \, ,
\end{equation}
where we have changed the upper limit of the sum from $m$, which is the number of terms in the Hamiltonian $\hat H(t)$ in Eq.~\eqref{chap:decoupling:eq:hamiltonian:sum}, to $n$, which is the dimension of the Lie algebra. We are allowed to do so because we can always set the coefficients $G_i(t) \equiv 0$ for any $i \geq m$ to recover the Hamiltonian.

We then consider the ansatz in Eq.~\eqref{chap:decoupling:time:evolution:ansatz}, which states that $\hat U(t)$ can be written as a product of operators $\hat U_j(t)$. When we then differentiate $\hat U(t)$ with respect to time $t$, we find the expression
\begin{align} \label{eq:decoupling:eq:differentiated:U}
\frac{d \hat U(t)}{dt} =  - i \, \sum_{j = 1}^n \dot{F}_j (t) \left( \prod_{k = 1}^{j - 1} \exp[- i \, F_k\, \hat H_k] \right) \, \hat H_j \, \left( \prod_{k = j}^n \exp[ - i \, F_k \, \hat H_k] \right)  ,
\end{align}
We then use the fact that $d \hat U(t) /dt = - i \hat H(t) \, \hat U(t)$ (which holds even when $\hat U(t)$ requires time-ordering). We then multiply Eq.~\eqref{eq:decoupling:eq:differentiated:U} by the inverse operator $\hat U^{-1}(t)$ on the right-hand-side and set the expression equal to $\hat H(t)$ in Eq.~\eqref{chap:decoupling:eq:hamiltonian:sum} to find
\begin{align} \label{eq:decoupling:eq:part:solved}
\sum_{j = 1}^n G_j(t) \, \hat H_j &= - i \,  \sum_{j = 1}^n \dot{F}_j (t) \left( \prod_{k = 1}^{j - 1} \exp[ - i \, F_k \, \hat H_k ] \right) \, \hat H_j \, \left( \prod_{k = j - 1}^1 \exp[ i \, F_k \, \hat H_k ] \right) \nonumber \\
&= - i \,  \sum_{j = 1}^n \dot{F}_j (t) \left( \prod_{k = 1}^{j - 1} \exp[ - i \, F_k \, ad\hat H_k ] \right) \, \hat H_j ,
\end{align}
where we have used the Baker-Campbell-Hausdorff Lemma (Lemma 1) in the second line.

By then applying Lemma~2 to the last line of Eq.~\ref{eq:decoupling:eq:part:solved}, we find
\begin{equation} \label{chap:decoupling:eq:differential:equation:sum}
\sum_{k = 1}^n G_k (t) \, \hat H_k =  - i \, \sum_{j = 1}^n \sum_{k = 1}^n \dot{F}_j (t) \, \xi_{kj} \, \hat H_k ,
\end{equation}
where we recall that $\xi_{kj}$ are analytic functions of the Hamiltonian parameters $G_j(t)$.

We note now that the operators $\hat H_k$ are linearly independent and effectively form a basis as part of the Lie algebra. We use the linear independence of $\hat H_k$ to find linear relations between the $G_k(t)$ functions and the $\dot{F}_j(t)$. They are related by the elements $\xi_{kj}$, which we can collect into a transformation matrix $\boldsymbol{\xi}$. We define the vector of Hamiltonian coefficients  $\vec{G} = (G_1, G_2, \cdots, G_n)^{\mathrm{T}}$ and the vector of $F$ coefficients $\dot{\vec{F}} = (\dot{F}_1 , \dot{F}_2 \,, \cdots \, \dot{F}_n)^{\rm{T}}$, such that
\begin{align} \label{eq:matrix:form:equation}
\vec{G} = - i \,  \boldsymbol{\xi}  \, \dot{\vec{F}}  .
\end{align}
Eq.~\eqref{eq:matrix:form:equation} encodes a number of $n$ differential equations and explicitly reads
\begin{align}
\begin{pmatrix}
G_1 \\ G_2 \\ \vdots \\ G_n
\end{pmatrix} = - i \,  \begin{pmatrix}
\xi_{11} & \xi_{12} & \cdots & \xi_{1n} \\
\xi_{21} & \xi_{22} &\cdots & \xi_{2n} \\
\vdots & \vdots & \ddots & \vdots \\
\xi_{n1} & \xi_{n2} & \cdots & \xi_{nn}
\end{pmatrix}
\begin{pmatrix}
\dot{F}_1 \\
\dot{F}_2 \\
\vdots \\
\dot{F}_n
\end{pmatrix}.
\end{align}
This system of differential equations can be solved if the matrix $\boldsymbol{\xi}$ is invertible. To determine if that is the case, we note that the determinant $\det{\boldsymbol{\xi}}$ is non-zero at $t = 0$, because $\boldsymbol{\xi}(0) = 1$. Thus, the matrix is invertible in some neighbourhood of $t = 0$. Since the matrix is invertible, we can write
\begin{equation}
\dot{\vec{F}} = f(\vec{G}, \vec{F}) = i \,  \boldsymbol{\xi}^{-1} \, \vec{G} .
\end{equation}
We now recall that we had the boundary condition $\vec{F}(0) = 0$, which ensures that $\hat U(t = 0) = \mathds{1}$. These boundary conditions ensure that there exists a unique solution to the system. However, the resulting differential equations may not always have analytic solutions, and therefore we might need numerical methods. This concludes the proof of the Decoupling Theorem. \qedsymbol

\subsection{A recipe for decoupling}
We here provide a summary of the decoupling methods in the form of a simple recipe that can be applied to any Hamiltonian that generates a finite Lie algebra.
\begin{enumerate}
\item Write the Hamiltonian $\hat H(t)$ as
\begin{equation} \label{chap:decoupling:eq:recipe:Hamiltonian}
\hat H(t) = \sum_j^m G_j(t) \hat H_j  ,
\end{equation}
and identify the functions $G(t)$ and the operators $\hat H_j$.
\item Identify the Lie algebra $L$ that is generated by the Hamiltonain by commuting the Hamiltonian terms $\hat H_j$  until they produce a set of $n$ operators that is closed under commutation. For example,  $[\hat H_1 , \hat H_2] \propto \hat H_3$.  Subsequently, $[\hat H_1, \hat H_3] \propto \hat H_4$, and so on, until all algebra elements are known.
\item State the ansatz for the time-evolution operator,
\begin{equation} \label{eq:recipe:ansatz}
\hat U(t) = \exp[ - i \, F_1(t) \, \hat H_1] \, \exp[ - i \, F_2(t) \, \hat H_2] \cdots \exp[ - i \, F_n (t) \, \hat H_n ] ,
\end{equation}
where the $F_j$-coefficients are dimensionless, time-dependent functions and the generators $\hat H_j$ are part of the Lie algebra $L$.
\item Differentiate the ansatz in Eq.~\eqref{eq:recipe:ansatz} with respect to time $t$ to find
\begin{align} \label{chap:decoupling:eq:recipe:differentiating}
\frac{d}{dt} \hat U(t) =& - i  \, \dot{F}_1 \, \hat H_1 \,  \hat{U}(t)  - i \, \dot{F}_2 \, \hat U_1 \hat H_2  \prod_{j = 2}^n \hat U_j  - i \, F_3 \, \hat U_1 \, \hat U_2 \, \hat H_3 \, \prod_{j = 3}^n \hat U_j  + \ldots - i \, \dot{F}_n \prod_{j = 1}^{n} \hat U_j \hat H_n  .
\end{align}
\item Multiply Eq.~\eqref{chap:decoupling:eq:recipe:differentiating} by $\hat U^{-1}(t)$ on the right and set the ansatz equal to the original Hamiltonian $\hat H(t)$:
\begin{equation} \label{eq:decoupling:congruence:multiplication}
\hat H(t) = \dot{F}_1 \, \hat H_1 + \dot{F}_2 \, \hat U_1 \, \hat H_2 \, \hat U_1^{-1} + \dot{F}_3 \, \hat U_2 \, \hat U_1 \, \hat H_3 \, \hat U_1^{-1} \, \hat U_2^{-1} + \ldots \, .
\end{equation}
\item Use the linear independence of $\{\hat H_j\}$ to construct a set of differential equations, where the solutions for $F_j$ depend on the original Hamiltonian coefficients $G_j(t)$. The equations are given by
\begin{equation} \label{chap:decoupling:eq:recipe:differential:equation:sum}
\sum_{k = 1}^n G_k (t) =  - i \, \sum_{j = 1}^n \sum_{k = 1}^n \dot{F}_j (t) \, \xi_{kj},
\end{equation}
where the functions $\xi_{kj}$ are obtained through the multiplications by congruence shown in Eq.~\eqref{eq:decoupling:congruence:multiplication}.
\item Solve the equations in Eq.~\eqref{chap:decoupling:eq:recipe:differential:equation:sum} analytically or numerically for the $F_j$ coefficients and use the result to determine the time-evolution operator $\hat U(t)$.
\end{enumerate}
There are many Hamiltonians that can be treated by using this recipe. In the next sections, we demonstrate how the dynamics of a harmonic oscillator Hamiltonian with linear and quadratic interaction terms can be solved.

\section{Hamiltonian with linear terms} \label{sec:linear:terms}
One of the simplest additions to a freely evolving quantum harmonic oscillator is a linear single-mode Hamiltonian interaction term. By linear, we mean that the term contains only single powers of operators. Such terms can correspond to a number of effects. Most commonly in optical systems, they represent continuous driving which arises, for example, by injecting laser light into a cavity~\cite{walls2007quantum}. If the laser light enters at a frequency different from the free frequency of the system, then the driving term change as a function of time.

In this section, we use the Lie algebra decoupling method to solve the dynamics of a single quantum harmonic oscillator with such linear driving terms.
The Hamiltonian that for such a system is given by
\begin{align} \label{eq:Hamiltonian:linear:driving}
\hat H(t) = \hbar \omega \hat a^\dag \hat a + \hbar \omega g_+ (t)  \hat a^\dag + \hbar \omega g_-(t)   \hat a ,
\end{align}
where $\omega$ is the free oscillation frequency and where $g_+(t)$ and $g_-(t)$ are complex, time-dependent and  dimensionless driving coefficients.

As a first step, we rescale time $t$ by the frequency $\omega$, such that $\omega t \rightarrow t$, where $t$ is now dimensionless.  Then, we start commuting the operators in the terms of Eq.~\eqref{eq:Hamiltonian:linear:driving}, using the commutator relations $[\hat a, \hat a^\dag ] = \mathds{1}$, $[ \hat a^\dag \hat a, \hat a] = - \hat a $, $[\hat a^\dag \hat a , \hat a^\dag] = \hat a^\dag$.
As a result, we see that the algebra that generates the evolution of this Hamiltonian is given by
\begin{align} \label{eq:linear:algebra}
\hat a^\dag \hat a,  && \hat a , &&  \hat a^\dag, && \mathds{1}, 
\end{align}
where $\mathds{1}$ is the identity operator.

We  then state the ansatz for the evolution operator $\hat U(t)$. It reads
\begin{align} \label{eq:linear:ansatz}
\hat U(t) = e^{- i F_0 \hat a^\dag \hat a} \, e^{- i F_+ \hat a^\dag} \, e^{- i F_- \hat a},
\end{align}
where we have introduced the coefficients $F_0, F_\pm \in \mathbb{C}$ as dimensionless functions of time. Here, we note that while the identity operator $\mathds{1}$ is part of the full algebra shown in Eq.~\eqref{eq:linear:algebra}, it results in a global phase when included in the ansatz in Eq.~\eqref{eq:linear:ansatz}. Such global phases can always be ignored, which is why we do not include $\mathds{1}$ in the ansatz. If instead we were to consider non-unitary dynamics, the contribution from the identity cannot be so easily dismissed (see Section~\ref{sec:open:dynamics}, where we discuss an extension of these methods to non-unitary dynamics). 

We now wish to use the Lie algebra decoupling method (see Section~\ref{sec:decoupling}) to derive the coefficients $F_0$ and $F_\pm$. We start by differentiating Eq.~\eqref{eq:linear:ansatz} with respect to time $t$. We then multiply the result by $\hat U^{-1}(t)$ on the right and compute the multiplications by congruence. Finally, we set the expression equal to the Hamiltonian in Eq.~\eqref{eq:Hamiltonian:linear:driving}. The details of the calculation can be found in~\ref{app:linear:terms:calculation}. 

We find that the coefficients $F_0$ and $F_\pm$ in the ansatz in Eq.~\eqref{eq:linear:ansatz} are given by 
\begin{align} \label{eq:linear:F:coefficients}
&F_0 = t ,
 &&F_+= \int^t_0 \mathrm{d}t^\prime \, g_+(t') \, e^{i t'},  &&
F_- =\int^t_0 \mathrm{d}t^\prime \, g_-(t') \, e^{- i t'} .
\end{align}
To evaluate the integrals in Eq.~\eqref{eq:linear:F:coefficients}, we must first choose a specific form of the driving functions $g_\pm(t)$. Once we have done so, we can fully fully characterise the system.

Let us now consider some examples of $g_\pm(t)$ and the dynamics that are generated by these choices. For the calculation that follows, we assume that the initial state of the system is a coherent state $\ket{\alpha}$, such that $\hat a \ket{\alpha} = \alpha \ket{\alpha}$ where $\alpha \in \mathbb{C}$.

We start by considering a constant coupling $g_\pm(t) \equiv g_0 $. For such a choice, the integrals in Eq.~\eqref{eq:linear:F:coefficients} evaluate to
\begin{align} \label{eq:linear:coefficients:constant}
&F_+ =  g_0 \left[ i - i \cos(t) + \sin(t) \right],
&&F_- = g_0 \left[ - i + i \cos(t) + \sin(t) \right].
\end{align}
We note that both $F_+$ and $F_-$ oscillate in time. 

\begin{figure*}
\centering
\subfloat[ \label{fig:linear:constant}]{
  \includegraphics[width=0.3\linewidth]{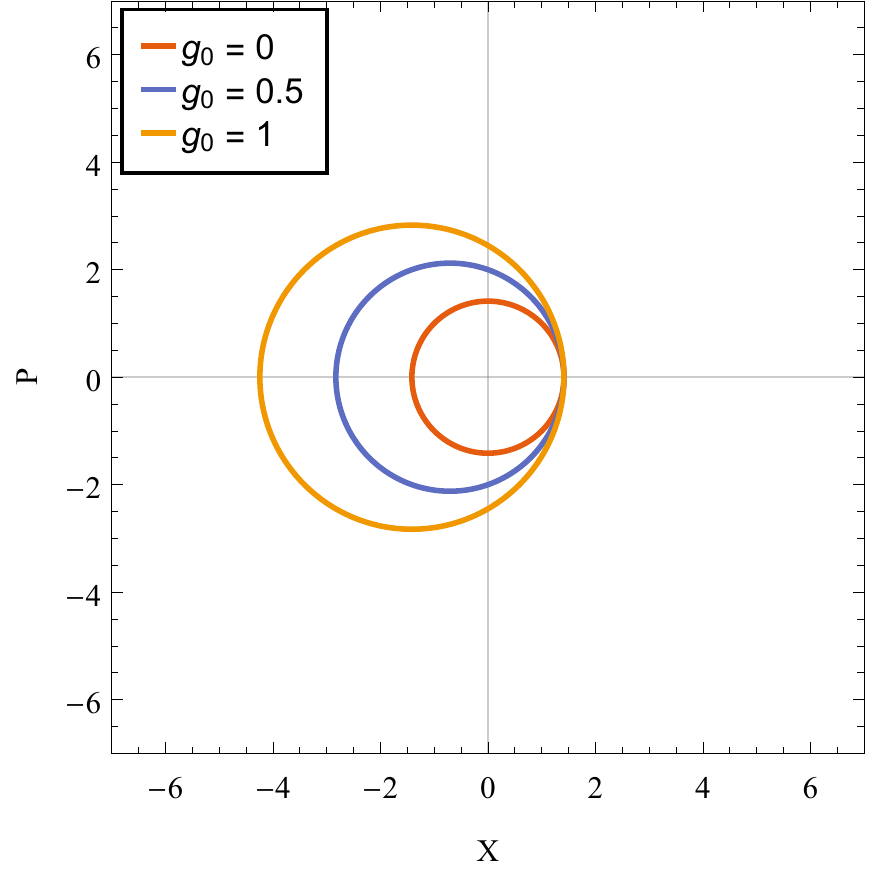}
}
\subfloat[ \label{fig:linear:resonant}]{
  \includegraphics[width=0.3\linewidth]{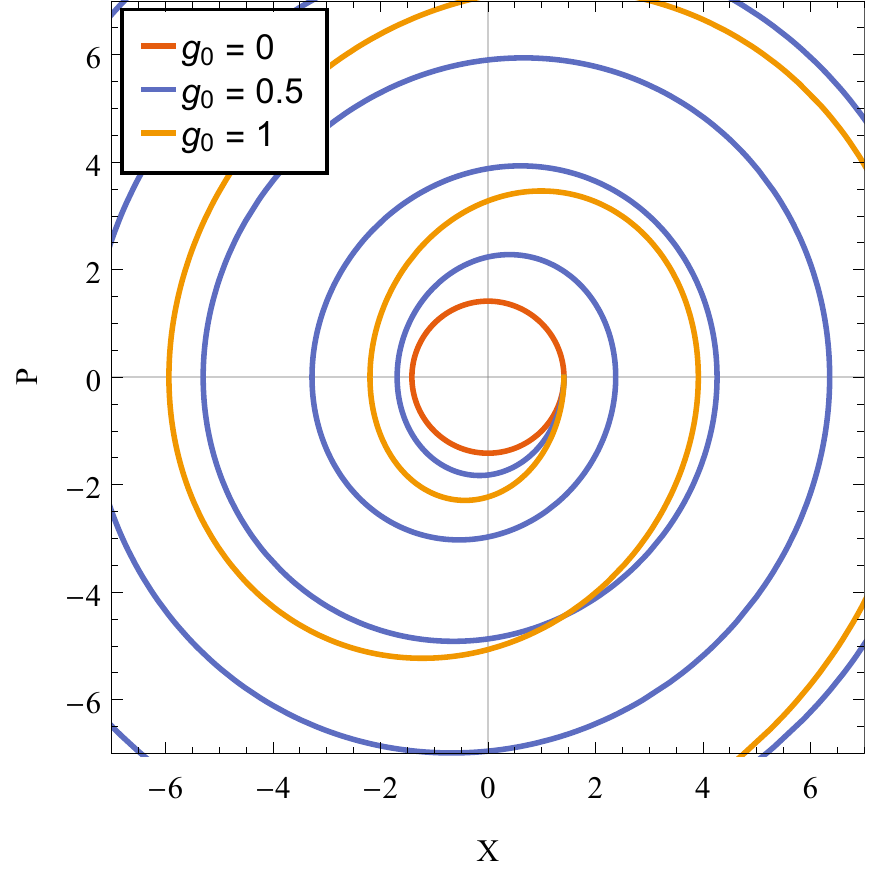}
}
\caption{ \textbf{Comparison between a Hamiltonian with constant and resonant linear terms}. Both plots show the quadratures $\hat X(t)$ and $\hat P(t)$ as a function of time for a constant linear term and a time-dependent linear term. When there is no driving ($g_\pm = 0$), the state explores a circle in phase space. Plot (a) shows the quadrature trajectories for a constant term $g_\pm \equiv g_0$. The state explores a limited trajectory in phase space. Plot (b) shows a time-dependent term with $g_\pm \equiv g_0 \cos(t + \phi)$, where different driving frequencies cause the state to explore wider and wider spirals in phase space, for $\phi = 0$ in this case. Both plots use the coherent state parameter $\alpha = 1$.  }
\label{fig:linear:drive}
\end{figure*}

To gain some intuition for the system dynamics, we consider the evolution of the position $\hat X$ and momentum $\hat P$ phase space quadratures. They are given by, in terms of the annihilation and creation operators,
\begin{align}
&\hat X = \frac{\hat a^\dag + \hat a }{\sqrt{2}}, && \mathrm{and} &&& \hat P = i \frac{\hat a^\dag - \hat a }{\sqrt{2}}.
\end{align}
In the Heisenberg picture, these quadratures evolve as $\hat X(t) = \hat U^\dag(t) \, \hat X \, \hat U(t)$ and $\hat P(t) = \hat U^\dag(t) \, \hat P \, \hat U(t)$. 
Given an initially coherent state $\ket{\alpha}$, the expectation values of the quadratures evolve as (see~\ref{app:linear:terms:calculation} for details of the derivation):
\begin{equation}
\begin{split} \label{eq:computed:quadratures}
\braket{\hat X(t)} &= \bra{\alpha} \hat U^\dag(t) \, \hat X \, \hat U(t) \ket{\alpha} =  \frac{1}{\sqrt{2}} \left[  e^{ i F_0 } \left( \alpha^*  + i F_-\right) + e^{- i F_0} \left( \alpha - i F_+ \right) \right],  \\
\braket{\hat P(t)} &=\bra{\alpha} \hat U^\dag(t) \, \hat P \, \hat U(t) \ket{\alpha} =  \frac{i}{\sqrt{2}} \left[  e^{ i F_0 } \left( \alpha^* + i F_-\right) - e^{- i F_0} \left( \alpha  - i F_+ \right) \right].
\end{split}
\end{equation}
We note that both $\braket{\hat X(t)}$ and $\braket{\hat P(t)}$ pick up time-dependent shifts that depend on the linear coupling strength $g_0$. Both quadratures return to their original point in phase space whenever $t$ is a multiple of $2\pi$. To better see this, we have plotted $\langle\hat X(t)\rangle$ and $\langle \hat P(t) \rangle$ in Figure~\ref{fig:linear:drive} as a function of time $t$ for  $\alpha = 1$. We note that as the value of  $g_0$ increases, the system performs larger  and larger trajectories in phase space.

If instead the functions $g_\pm(t)$ change as functions of time, the system behaviour becomes much more involved. Here, we find that interesting effects such as resonances markedly affect the dynamics of the system. By resonance, we refer to time-dependent effects that occur at a frequency equal to the free frequency $\omega$.

To explore the resonant case, we let the functions $g_\pm(t)$ both oscillate in time with $g_\pm(t) = g_0 \cos(t + \phi)$, where $g_0$ is again the amplitude and $\phi$ is a phase offset. By solving the integrals in Eq.~\eqref{eq:linear:F:coefficients}, we find that the coefficients $F_\pm$ become
\begin{align} \label{eq:linear:resonant:drive:F:coefficients}
&F_{+} = \frac{g_0}{2}\left[ t \, e^{- i \phi} +e^{i \phi} \, e^{i t}  \sin(t)  \right],
&&F_{-} =\frac{g_0}{4} \left[ i e^{- i (2 t + \phi)}  - i e^{- i \phi}  + 2 t \right].
\end{align}
We note that, compared with the coefficients in Eq.~\eqref{eq:linear:coefficients:constant}, which arise for constant coupling, both $F_+$ and $F_-$ in Eq.~\eqref{eq:linear:resonant:drive:F:coefficients} now increase linearly in time. As a result, the state does not return to its original position in phase space.

We again plot $\braket{\hat X(t)}$ and $\braket{\hat P(t)}$ for the resonant driving as a function of time $t$.  The result can be found  in Figure~\ref{fig:linear:resonant} for different driving strengths $g_0$ and the phase choice $\phi = 0$. We note that as $t$ increases, the state is exploring larger and larger trajectories in phase space.

\section{Hamiltonian with quadratic terms} \label{sec:quadratic:drive}

We now consider harmonically trapped systems with additional quadratic Hamiltonian terms. Such terms can be engineered by for example changing the trapping frequency of the system~\cite{gieseler2012subkelvin}.
In cases where the quadratic terms are modulated at twice the free frequency, the term is known as a parametric drive~\cite{kinsler1991quantum}. In fact, modulating the potential at parametric resonance for a specific phase offset causes a reduction in the number of quanta in a harmonic oscillator~\cite{kinsler1991quantum, manikandan2022cooling}. In addition, parametric modulations can in certain cases enhance the sensitivity of a quantum force sensor~\cite{qvarfort2021optimal}.

The Hamiltonian for a quantum harmonic oscillator with quadratic single-mode interaction terms reads
\begin{align} \label{eq:quantum:harmonic:oscillator}
\hat H(t) = \hbar \omega \hat a^\dag \hat a + \hbar \omega \lambda_+ (t)  \hat a^{\dag2 } + \hbar \omega \lambda_-(t) \hat a^2 ,
\end{align}
where $\omega$ is the free oscillation frequency of the mode, and where $\lambda_\pm(t)$ are complex, time-dependent and dimensionless coefficients.

To solve the dynamics induced by the Hamiltonian in Eq.~\eqref{eq:quantum:harmonic:oscillator}, we start by defining the following quadratic operators
\begin{align} \label{eq:K:algebra}
&\hat K_+ = \frac{1}{2}\hat a^{\dag 2} , && \hat K_0 = \frac{1}{4} \left( 2 \hat a^\dag \hat a + \mathds{1} \right), && \hat K_- = \frac{1}{2} \hat a^2 .
\end{align}
These operators form an SU(1,1) algebra and obey the following commutation relations
\begin{align}
&[\hat K_0 , \hat K_{\pm}] = \pm \hat K_{\pm}, &&[\hat K_+ , \hat K_-] = - 2 \hat K_0.
\end{align}
This SU(1,1) algebra shares many properties with the commonly used SU(2) algebra, which induces the dynamics of two-level systems, for example. If SU(2) can be thought of as a sphere, the SU(1,1) algebra instead represents the two semi-spheres that make up the full sphere. For more background on the applications of SU(1,1) in quantum physics, see Ref~\cite{chiribella2006applications}.

We proceed to state the following ansatz for the time-evolution generated by the Hamiltonian in Eq.~\eqref{eq:quantum:harmonic:oscillator}:
\begin{align} \label{eq:quadratic:ansatz}
\hat U(t) = e^{- i  \xi_+ \hat K_+}  \, e^{ - i\xi_0 \hat K_0} \, e^{- i \xi_- \hat K_-},
\end{align}
where $\xi_\pm$ and $\xi_0$ are complex time-dependent coefficients that we wish to solve for. 

We note that we can reorder the exponentials in Eq.~\eqref{eq:quadratic:ansatz} into an expression that includes a single-mode squeezing operator and a rotation with $\hat a^\dag \hat a$.  A single-mode squeezing operator is defined as $\hat{S}(\zeta) =e^{(\zeta \hat{a}^{\dagger 2} - \zeta^* \hat{a}^2)/2}$, with the complex parameter $\zeta=r e^{i \varphi}$ that includes the strength and phase-space direction of squeezing.

As in the previous section, we follow the decoupling method outlined in Section~\ref{sec:lie:algebra} to derive the differential equations for $\xi_0$ and $\xi_\pm$. We start by differentiating $\hat U(t)$ in Eq.~\eqref{eq:quadratic:ansatz} with respect to time $t$. We then multiply the result by $\hat U^{-1}(t)$ on the right and compute the resulting congruence relations. We then set the result equal to the Hamiltonian in Eq.~\eqref{eq:quantum:harmonic:oscillator} and use the linear independence of the operators to find the differential equations
See~\ref{app:quadratic:terms:calculation} for details of this calculation.

We find the following differential equations for $\xi_\pm$ and $\xi_0$:
\begin{align} \label{eq:quadratic:diff:eqs}
\dot{\xi}_+& = \lambda_+(t)  + \omega \,\xi_+ + \lambda_-(t) \, \xi_+^2,  \nonumber \\
\dot{\xi}_0    &=  \omega  +2 \lambda_-(t)\, \xi_+,   \\
\dot{\xi}_-&=  \lambda_-(t) \,  e^{i \xi_0}. \nonumber
\end{align}
We note that only the first and second differential equations in Eq.~\eqref{eq:quadratic:diff:eqs} are coupled. This is because the time-evolution operator in Eq.~\eqref{eq:quadratic:ansatz} corresponds to a Bogoliubov transformation, which can be fully characterised by two free parameters. As a result, the coefficient $\xi_-$ can be fully determined once the first differential equation has been solved.

When the amplitudes $\lambda_\pm(t)$  are constant in time, such that $\lambda_\pm(t) \equiv \lambda_\pm$, the differential equations in Eq.~\eqref{eq:quadratic:diff:eqs} can be solved to give
\begin{align} \label{eq:quadratic:algebra:constant:solutions}
\xi_{\pm} & =  \frac{\lambda_\pm}{ \Gamma}   \left[   \frac{  \sinh( t \Gamma)}{\cosh(t \Gamma) + \frac{i}{2\Gamma}  \sinh( t \Gamma) } \right] ,\nonumber \\
\xi_0 & = - 2  i \ln\left[\cosh(\Gamma) + \frac{i}{2 \Gamma} \sinh(\Gamma)\right],
\end{align}
where we have defined the parameter $\Gamma^2  =\lambda_+ \lambda_- -  \frac{1}{4}$.
As a result, if $\lambda_\pm$ are constant, we do not need to consider time-ordering. We can write down the following simple relation:
\begin{align}  \label{eq:quadratic:decoupling:constant}
e^{ - i  t  \left( \lambda_+ \hat K_+ + \omega \hat K_0 + \lambda_- \hat K_- \right)} = e^{- i  \xi_+ \, \hat K_+}  \, e^{- i\xi_0 \, \hat K_0} \, e^{ - i \xi_- \, \hat K_-}, 
\end{align}
which is well-known in quantum optics~\cite{barnett2002methods}.

Our method presented here, however, directly generalizes to time-dependent coefficients. In this case, and depending on the form of the time-dependence, the differential equations in Eq.~\eqref{eq:quadratic:diff:eqs} must usually be solved numerically. There are however cases where they reduce to well-known differential equations, such as the Mathieu equation~\cite{qvarfort2020time, manikandan2022cooling}.

Before moving on, we note that the quadratic Hamiltonian in Eq.~\eqref{eq:quantum:harmonic:oscillator} can be cast as a Hamiltonian matrix and solved using a phase space Lie algebra decoupling\footnote{For more information about phase space methods for continuous variable quantum systems, see Ref~\cite{serafini2017quantum}}. We outline this decoupling methods in~\ref{sec:phase:space}. The mapping between the phase space solution and the Hilbert space solution is usually non-trivial. Sometimes, however, the phase space solution may yield differential equations that are simpler to solve, compared with the Hilbert space solution. It is generally difficult to say in advance which solution is the easiest to work with.

\section{Most general Gaussian Hamiltonian} \label{sec:linear:quadratic:drive}
Once we have identified two closed algebras, we can combine them to obtain more general solutions, provided that the full algebra remains finite. For example, we can combine the solutions for the linear and quadratic driving terms that we derived in Sections~\ref{sec:linear:terms} and~\ref{sec:quadratic:drive}. Doing so provides us with a solution for the most general dynamics of a harmonic oscillator with Gaussian terms\footnote{The use of ‘Gaussian’ here refers to the fact that Hamiltonians with linear and quadratic terms map Gaussian states to Gaussian other Gaussian states. To map a Gaussian state to a non-Gaussian state, we instead need a Hamiltonian with cubic or higher-order terms.}. If the coefficients are constant in time, we can usually compute them directly, as for example in Ref~\cite{pikovski2014macroscopic}. In general though, the coefficients may be time-dependent.  

To solve the dynamics of a Hamiltonian with both linear and quadratic interaction terms, we could start by writing down the full algebra and follow the decoupling recipe in Section~\ref{sec:decoupling}. The full algebra has five unique elements, so this  necessarily involves multiplying  out each of the terms in the ansatz and solving a set of five simultaneous differential equation. However, we can instead make use the fact that we already know the solution for one of the subalgebras (e.g. the quadratic one in Section~\ref{sec:quadratic:drive}). By then considering an interaction picture that rotates with one of the subalgebras, we are able to determine the effects of the solutions onto the second subalgebra. In this way, it is possible to partition the dynamics and solve the individual contributions separately. It should be noted that there is no right or unique way to perform this partition, and sometimes one partition works better than the other.

We start by combining the Hamiltonians in Eq.~\eqref{eq:Hamiltonian:linear:driving} and Eq.~\eqref{eq:quantum:harmonic:oscillator} into a single Hamiltonian with both linear and quadratic terms:
\begin{align} \label{eq:linear:quadratic:Hamiltonian}
\hat H (t) = \hat H_0  + \hat H_{L}(t) + \hat H_{Q}(t) ,
\end{align}
where  we have used the subscripts $L$ and $Q$ to denote the linear and quadratic terms, respectively. The Hamiltonian contributions in Eq.~\eqref{eq:linear:quadratic:Hamiltonian} are given by 
\begin{align}
\hat H_0 &= \hbar \omega \hat a^\dag \hat a,  \nonumber \\
\hat H_{L}(t) &=  \hbar \omega g_+(t) \hat a^\dag + \hbar \omega g_-(t) \hat a,   \\
\hat H_{Q}(t) &=  \hbar \omega \lambda_+(t) \hat a^{\dag2} + \hbar \omega \lambda_- (t) \hat a^2,\nonumber
\end{align}
As before, $\omega$ is the  angular frequency of the free mode, $g_\pm(t)$ are the coefficients of the linear terms, and $\lambda_\pm(t)$ are the coefficients of the quadratic terms.

The full algebra generated by the Hamiltonian in Eq.~\eqref{eq:linear:quadratic:Hamiltonian} is now given by
\begin{align}
& \hat a ,&& \hat a^\dag ,&& \mathds{1}, \nonumber \\
&\hat K_0, && \hat K_+, && \hat K_-,
\end{align}
with $\hat K_0$ and $\hat K_\pm$ defined in Eq.~\eqref{eq:K:algebra}. We can check that combining the algebras in this way does not generate any new operators by examining the new commutator relations:
\begin{align} \label{eq:invariant:algebra}
&[\hat K_0, \hat a  ] =-  \frac{1}{2} \hat a, && [\hat K_+, \hat a] = - \frac{1}{2} \hat a^\dag,   \\
&[\hat K_0 , \hat a^\dag] = \frac{1}{2} \hat a^\dag, && [\hat K_-, \hat a^\dag] = \frac{1}{2} \hat a .
\end{align}
From this, we see that commuting the linear operators with the quadratic operators leaves the linear algebra invariant and does not add any new operators to the full algebra.

We could now proceed to solve the system for the full algebra in Eq.~\eqref{eq:invariant:algebra}. It means we would have to compute all five multiplications by congruence according to Eq.~\eqref{eq:decoupling:congruence:multiplication} and then solve the resulting differential equations. While this is certainly possible (and might in this case not be too challenging), there is, as mentioned at the beginning of this section, an easier alternative, which involves defining two rotating frames defined by the two subalgebras. This is similar to the notion of solving the Schrödinger equation in the interaction picture.

There is no set recipe for how to best partition the dynamics, but often one composition is easier to treat than the other. In our case, we choose to focus on the linear terms and how they evolve under the quadratic subalgebra. The reason for this choice is that the quadratic algebra leaves the linear algebra invariant, while the action of the linear algebra of the quadratic one reintroduces linear components into the ansatz. We consider a frame that rotates with the quadratic Hamiltonian $\hat H_{Q}(t) $. The time-evolution generated by the quadratic part of Hamiltonian is given by
\begin{align} \label{eq:UQ}
\hat U_{Q}(t) = \mathcal{T} \mathrm{exp} \left[ - \frac{i}{\hbar} \int^t_0 \mathrm{d}t’ \, \left( \hat H_0 + \hat H_{Q}(t’) \right) \right].
\end{align}
Note that we have included the free evolution $\hat H_0$ in Eq.~\eqref{eq:UQ} to complete the quadratic algebra. We already know the solution to $\hat U_{Q}(t)$, with the ansatz shown in Eq.~\eqref{eq:quadratic:ansatz}, and the differential equations for the coefficients listed in Eq.~\eqref{eq:quadratic:diff:eqs}.

Next, we consider how the Hamiltonian $\hat H_{L}(t)$ evolves in the frame rotating with $\hat U_{Q}(t)$. Applying this solution to $\hat U_Q(t)$ in Eq.~\eqref{eq:quadratic:ansatz} to $\hat H_L(t)$, we find that it evolves as (see~\ref{app:linear:quadratic:terms} for the calculation details):
\begin{align} \label{eq:evolving:Hamiltonian}
 \hat U_{Q}^\dag(t) \, \hat H_{L}(t) \, \hat U_{Q}(t)
&= \hbar \omega \left[ \mu(t) \hat a  + \nu (t) \hat a ^\dag \right],
\end{align}
where the two coefficients $\mu(t)$ and $\nu(t)$ are given by
\begin{equation}
\begin{split}
\mu(t) &= g_+(t) e^{\frac{1}{2} i \xi_0}  - i g_-(t) \xi_+ e^{\frac{1}{2} i \xi_0},   \\
\nu (t) &=  i g_+(t) \xi_-  e^{\frac{1}{2} i \xi_0}  + g_-(t) \left( e^{- \frac{1}{2} i \xi_0} + \xi_+ \xi_- e^{\frac{1}{2} i \xi_0 } \right) .
\end{split}
\end{equation}
The full time-evolution of the linear subalgebra in the rotating frame is then 
\begin{align}
\hat U_{L}(t) = \mathcal{T} \mathrm{exp}\left\{ - \frac{i}{\hbar} \int^t_0 \mathrm{d}t’ \, \left[ \mu(t') \hat a + \nu(t') \hat a^\dag \right] \right\}.
\end{align}
Returning to the lab frame, as per the standard interaction picture treatment, we multiply $\hat U_L(t)$ by the left such that the full solution in the Schrödinger equation then reads
\begin{align}
\hat U(t) = \hat U_{Q}(t) \, \hat U_{L}(t).
\end{align}
We can find the solution for $\hat U_{L}(t)$ by using the same recipe as in Section~\ref{sec:linear:terms}. We know that the algebra operators of the linear algebra are given by
\begin{align}
& \hat a ,&& \hat a^\dag ,&& \mathds{1},
\end{align}
and as before, we ignore the identity operator, since it imparts a global phase. Note that here we have not included the free evolution $\hat a^\dag \hat a$, since that has already been included in the quadratic algebra.
Just like in Section~\ref{sec:linear:terms}, we make the ansatz:
\begin{align} \label{eq:linear:quadratic:ansatz}
\hat U_{L}(t) = e^{- i F_+ \hat a^\dag} \, e^{- i F_- \hat a},
\end{align}
where the $F$-coefficients are now different from that in Eq.~\eqref{eq:linear:F:coefficients}, because the Hamiltonian is given by that in Eq.~\eqref{eq:evolving:Hamiltonian}. The solutions to $F_\pm$ in Eq.~\eqref{eq:linear:quadratic:ansatz} are given by
\begin{align}
&F_+ = \int^t_0 \mathrm{d}t^\prime \, \nu(t') ,  &&F_- =\int^t_0 \mathrm{d}t^\prime \, \mu (t') .
\end{align}
These integrals might be challenging to perform analytically, depending on the explicit time-dependence of the coefficients, which in turn may depend on the pulse shape of a driving optical field, for example.  The above expressions do however capture the dynamics exactly.

We are then finally in a position to write down the full solution to the dynamics generated by the Hamiltonian in Eq.~\eqref{eq:linear:quadratic:Hamiltonian}. It is given by
\begin{align}
\hat U(t) =  e^{-i \xi_+ \hat K_+}  \, e^{-i\xi_0 \hat K_0} \, e^{-i \xi_- \hat K_-} \, e^{-i F_+ \hat a^\dag} \, e^{-i F_- \hat a}.
\end{align}
This procedure, where two subalgebras are combined into a single closed algebra, can in principle be repeated for more than one mode, as long as the complete algebra remains finite.

\section{Discussion} \label{sec:discussion}
We have solved the dynamics of three different Hamiltonians with linear, quadratic, and a combination of linear and quadratic interaction terms, respectively.  While this class of Hamiltonians remains Gaussian, the results still capture a wide range of physical situations, especially in quantum optics~\cite{barnett2002methods}. 
There are a few additional mathematical aspects of the Lie algebra decoupling method that are worth mentioning, such as the notion of symmetries, also known as Casimir elements, and the extension of the decoupling method to open-system dynamics.

\subsection{Casimir elements}

A Casimir element is an element that is part of the algebra, but which commutes with all other operators in the algebra. A trivial example is the identity operator $\mathds{1}$, which arises as part of the linear algebra shown in Eq.~\eqref{eq:linear:algebra} from when we take the following commutator: $[\hat a^\dag + \hat a , i( \hat a^\dag - \hat a) ] = 2 i \mathds{1}$. Since the identity operator commutes with every other element in the algebra, it can be considered a Casimir element.

The Casimir element can however also be non-trivial. For example, in nonlinear cavity optomechanics~\cite{aspelmeyer2014cavity}, where a mechanical oscillator is coupled through radiation-pressure to an optical mode, the Hamiltonian is given by
\begin{align} \label{eq:OMS:Hamiltonian}
\hat H_{OMS} = \hbar \omega_c \hat a^\dag \hat a + \hbar \omega_m \hat b^\dag\hat b - \hbar g \hat a^\dag \hat a (\hat b^\dag + \hat b ), 
\end{align}
where $\omega_c$ is the optical oscillation frequency of the optical mode with annihilation and creation operators $\hat a, \hat a^\dag$,  $\omega_m$ is the mechanical oscillation frequency of the mechanical mode with annihilation and creation operators $\hat b, \hat b^\dag$, and $g$ is the coupling strength between the optical and mechanical modes. The dynamics of this Hamiltonian with additional linear and quadratic mechanical terms has been previously solved in full generality~\cite{qvarfort2019enhanced, qvarfort2020time}. 

We note that this Hamiltonian is the same as the linearly driven quantum harmonic oscillator explored in Section~\ref{sec:linear:terms}, except that it is now multiplied with the operator $\hat a^\dag \hat a$. If we proceed to map out the algebra of this system by taking the commutator between the free evolution of the mechanical mode and the interaction term in Eq.~\eqref{eq:OMS:Hamiltonian}, we find the following term:
\begin{align}
[\hat b^\dag \hat b, \hat a^\dag \hat a (\hat b^\dag + \hat a ) ] = \hat a^\dag \hat a (\hat b^\dag - \hat b).
\end{align}
Then, commuting this new term with the original interaction term, we find
\begin{align}
[\hat a^\dag \hat a (\hat b^\dag + \hat b), \hat a^ \dag \hat a( \hat b^\dag - \hat b)] = 2 (\hat a^\dag \hat a )^2.
\end{align}
The operator $(\hat a^\dag \hat a)^2$ corresponds to a new algebra element. It is a self-Kerr interaction that can also be identified by diagonalising the Hamiltonian with a polaron transform~\cite{bose1997preparation, mancini1997ponderomotive}. We note that $(\hat a^\dag \hat a)^2$ commutes with all existing algebra elements, which means that it is a Casimir element. This time, however, we cannot discard it since acting on a state leads to observable changes in the quantum state. We also note that the nonlinear optomechanical Hamiltonian is one of the few examples of a non-quadratic Hamiltonian with a closed algebra, which allows for the dynamics to be solved exactly.

\subsection{Extension to open dynamics} \label{sec:open:dynamics}
Thus far, we have exclusively focused on quantum systems isolated from their environment, which results in unitary (closed) dynamics. However, in reality, every quantum system in the laboratory interacts with its surrounding. In cavity QED, for example, such interactions can take the form of light leaking out from a cavity, which leads to dissipation~\cite{walls2007quantum}. Dephasing noise is another ubiquitous source of decoherence, such as in superconductive qubits~\cite{krantz2019quantum}. Quantum systems may also be subject to dissipation and thermalization due to quantum Brownian motion~\cite{breuer2002theory}.

One of the most common ways in which open dynamics is modelled is with the Gorini-Kossakowski-Sudarshan-Lindblad equation~\cite{gardiner2004quantum}, also known as just the Lindblad equation. It is the most general Markovian master equation and is given by
\begin{equation} \label{eq:Lindblad:general}
\dot{\hat{\varrho}} = - i [\hat H, \hat \varrho] + \sum_{n,m=1}^{N^2-1} h_{nm} \left(  \hat L_n  \, \hat \varrho \, \hat L_m^\dagger - \frac{1}{2} \{ \hat L_m^\dagger \hat L_n , \hat \varrho \} \right),
\end{equation}
where $\hat \varrho$ is the density matrix of a quantum state, $\hat H$ is the Hamiltonian operator,  $\hat L_n$ is a phenomenological Lindblad operator that captures the effect of the surrounding environment, and where  $\{\cdot, \cdot \}$ denotes the anti-commutator.

If we now write the state $\hat \varrho(t)$ as a vector, which we can do by stacking either its rows or columns, such that  $\hat \varrho \rightarrow \kket{\varrho}$, we can rewrite Eq.~\eqref{eq:Lindblad:general} as a matrix equation
\begin{align} \label{eq:vectorised:Lindblad}
\frac{d}{dt} \kket{\varrho}   = \hat{\mathcal{L}}\kket{\varrho},
\end{align}
The Lindbladian  $\hat{\mathcal{L}}$ is now given by
\begin{align} \label{eq:LH:LL:definitions}
\hat{\mathcal{L}}(t) &= - i \bigl[ \hat H(t) \otimes \mathds{1} -  \mathds{1}\otimes \hat H^{\rm{T}}(t) \bigr]  +   \sum_{n,m = 1}^{N^2-1} \frac{h_{nm}}{2} \left[ 2 \hat L_n\otimes \hat L_m^{\dagger \rm{T}}- \hat L_m^\dagger \hat L_n \otimes \mathds{1} + \mathds{1}\otimes  (\hat L_m^\dagger \hat L_n)^{\rm{T}} \right],
\end{align}
where $\mathrm{T}$ indicates the transpose of an operator. This is also known as a Louville-space extension~\cite{gyamfi2020fundamentals}. The additional mode indicated with the tensor product in Eq.~\eqref{eq:LH:LL:definitions} appears due to the doubling of the Hilbert space, which is required in order to describe the density matrix as a vector rather than a matrix. To derive the expression in Eq.~\eqref{eq:LH:LL:definitions}, we have used the following identity which arises when vectorising products of operators: $\kket{\hat A \hat B \hat C} = \hat A \otimes \hat C^{\mathrm{T}} \kket{\hat B}$. The appearance of the transpose in these expressions can be eliminated by considering the operators in a basis with real matrix entries, such as the Fock basis.

The formal solution to Eq.~\eqref{eq:vectorised:Lindblad} is then given by
\begin{equation} \label{eq:formal:solution:S}
\hat{\mathcal{S}}(t) = \mathcal{T} \mathrm{exp} \left[ \int^t_0 \mathrm{d} t' \, \hat{\mathcal{L}}(t') \right],
\end{equation}
such that the initial density matrix $\kket{\varrho(t = 0)}$ evolves as $\kket{\varrho(t)} = \hat{\mathcal{S}}(t) \kket{\varrho(t = 0)}$.

The expression in Eq.~\eqref{eq:formal:solution:S} now looks similar to the definition of the unitary time-evolution operator in Eq.~\eqref{eq:U:formal:solution}. It turns out that $\hat{\mathcal{S}}(t)$ is also amenable to the Lie algebra decoupling method.
To see this, we consider the algebra that generates $\hat{\mathcal{S}}(t)$. It is effectively a two-mode algebra because of the vectorisation process. We can then generate the full algebra by commuting the following elements of $\hat{\mathcal{L}}$:
\begin{align} \label{eq:non:hermitian:algebra:elements}
&\hat H\otimes \mathds{1}, && \mathds{1}\otimes \hat H ,\nonumber \\
& \hat L_n \otimes \hat L_m^{\dag \mathrm{T}}, && \hat L^\dag_m \hat L_n \otimes \mathds{1}, && \mathds{1} \otimes (\hat L_m^\dag \hat L_n)^{\mathrm{T}}.
\end{align}
If this algebra is closed with $N$ elements, we may use the decoupling theorem in Section~\ref{sec:decoupling} to make an ansatz of the form
\begin{align}
\hat{\mathcal{S}}(t) = \hat{\mathcal{S}}_1(t) \, \hat{\mathcal{S}}_2(t) \ldots \hat{\mathcal{S}}_n(t),
\end{align}
where each $\hat{\mathcal{S}}_j(t)$ is now given by $\hat{\mathcal{S}}_j(t) = e^{ D_j \hat L_j}$, for which $D_j$ is a complex coefficient and $\hat L_j$ is one of the $n$ algebra elements generated from taking the commutators between the terms shown in in Eq.~\eqref{eq:non:hermitian:algebra:elements}.
By the same logic that all quadratic Hamiltonians generate a closed algebra, any quadratic Hamiltonian in combination with linear Lindblad operators should always generate a closed algebra.

Before moving on, we mention a few examples of treatments of open system dynamics with Lie algebras in the literature. The solution for a self-Kerr oscillator interacting with a thermal environment can be found in Ref~\cite{chaturvedi1991solution}, and a exploration of coupled harmonic oscillators can be found in Ref~\cite{teuber2020solving}. We also note that sometimes, such as in the case of an optomechanical system with optical dissipation, it is not necessary for the algebra to be closed in order to derive certain closed-form expressions of system quantities~\cite{qvarfort2021master}.

\section{Conclusions} \label{sec:conclusions}
In this work, we have provided a pedagogical introduction to solving the dynamics of quantum systems using a Lie algebra decoupling method. 
As a demonstration of the method, we considered a quantum harmonic oscillator with linear and quadratic time-dependent single-mode Hamiltonian interaction terms. For each example, we derived the exact differential equations that govern the dynamics. The result is a model for the most general Gaussian Hamiltonian with arbitrary time-dependent interaction terms. Such a model could potentially be applied to a number of problems in quantum optics and related fields, with applications for quantum technologies and fundamental physics. 
Finally, we note that the methods presented here apply to a number of Hamiltonians as long as they map to similar algebras, such as the two-mode beamsplitter interaction Hamiltonian. 

\section*{Acknowledgments}
We thank Yuefei Liu, Suocheng Zhao, Sreenath K.~Manikandan, and David Edward Bruschi for fruitful discussions and comments. S.Q.~is funded in part by the Wallenberg Initiative on Networks and Quantum Information (WINQ) and in part by the Marie Skłodowska--Curie Action IF programme \textit{Nonlinear optomechanics for verification, utility, and sensing} (NOVUS) -- Grant-Number 101027183. I.P. acknowledges support by the Swedish Research Council under grant no.~2019-05615, the European Research Council under grant no.~742104 and The Branco Weiss Fellowship -- Society in Science.

\appendix

\section{Lie algebra decoupling theorem in phase space} \label{sec:phase:space}
In the study of quantum continuous variables, it is widely known that Gaussian states are completely characterised by their first and second moments. As a result, it is possible to study the dynamics of Gaussian states by virtue of modelling the evolution of these moments alone. 
The Lie algebra decoupling theorem can be extended to phase space methods. We shall see that it is sometimes easier to solve the dynamics this way, as the problem of computing non-trivial commutators and multiplications by congruence is reduced to that of matrix multiplication. For an introduction to quantum continuous variables, see Ref~\cite{serafini2017quantum}.

\subsection{Link between phase-space and Hilbert space}
We start by defining a vector of  first moments $\hat{\mathbb{X}} = (\hat a_1 , \hat a_2, \hat a_3, \ldots \hat a_N, \hat a_1^\dag, \hat a_2^\dag , \hat a_3^\dag , \ldots, \hat a_N^\dag)^{\mathrm{T}}$, where $\hat a_1, \hat a_2, \hat a_3 \ldots \hat a_N$ are the annihilation operators for a number of $N$ modes. The evolution of these moments with respect to some Hamiltonian $\hat H(t)$ is then given by
\begin{equation}
\hat{\mathbb{X}}(t) = \mathbf{S}(t) \hat{\mathbb{X}}, 
\end{equation}
where $\mathbf{S}(t)$ is a symplectic matrix given by
\begin{equation}
\mathbf{S}(t) = \mathcal{T} \mathrm{exp}\left[ \mathbf{\Omega} \int^t_0 \mathrm{d}t' \, \mathbf{H}(t') \right].
\end{equation}
Here, $\mathcal{T}$ indicates time-ordering of the exponential,  $\mathbf{\Omega}$ is the symplectic form that encodes the commutator relations, and $\mathbf{H}(t)$ is the Hamiltonian matrix of the Hamiltonian operator $\hat H$, defined as
\begin{align}
\hat H(t) = \frac{1}{2} \hat{\mathbb{X}}^{\mathrm{T}} \, \mathbf{H}(t) \, \hat{\mathbb{X}}.
\end{align}
In the basis of the annihilation and creation operators, the symplectic form $\mathbf{\Omega}$ is given by
\begin{align}
&\mathbf{\Omega} = \bigoplus_{n = 1}^N \mathbf{\Omega}_1 , &&\mathrm{where} &&& \mathbf{\Omega}_1 = \begin{pmatrix} i & 0 \\ 0 & -i \end{pmatrix} ,
\end{align}
where $N$ is the number of modes under consideration.
In other bases, such as the position and momentum basis, $\mathbf{\Omega}$ is instead given by
\begin{align}
\mathbf{\Omega}_1 = \begin{pmatrix} 0 & 1 \\ 1 & 0 \end{pmatrix}  .
\end{align}
The reason that we in this work choose to work in the $\{\hat a, \hat a^\dag\}$ basis is because it is generally easier to commute the free evolution term $\hat a^\dag \hat a$ with the interaction terms in the Hamiltonian, and thereby predict the effects of time-evolution. 
The link between the dynamics in the Hilbert space and phase space is
\begin{align} \label{eq:hilbert:space:symplectic:link}
\hat U^\dag (t)\,\hat{\mathbb{X}} \,\hat U(t)  = \mathbf{S}(t) \hat{\mathbb{X}}. 
\end{align}
Just like with the Hilbert space method, we can make an ansatz for the solution of $\mathbf{S}(t)$:
\begin{align} \label{eq:phase:space:ansatz}
\mathbf{S}(t) = \mathbf{S}_1 (t) \, \mathbf{S}_2 (t) \, \mathbf{S}_3 (t) \ldots \mathbf{S}_n(t) = e^{F_1 \mathbf{\Omega} \mathbf{H}_1} \,e^{F_2 \mathbf{\Omega} \mathbf{H}_2} \, e^{F_3 \mathbf{\Omega} \mathbf{H}_3} \ldots e^{F_n \mathbf{\Omega} \mathbf{H}_n },
\end{align}
where $n$ is the number of elements in the Lie algebra. Note here that our Lie algebra does not just consist of the Hamiltonian matrices $\mathbf{H}_j$, but rather the product of these matrices with the symplectic form. That is, the algebra elements are given by $\mathbf{\Omega} \mathbf{H}_j$.
If we then consider the individual decoupled contributions to $\hat U(t)$, the relationship reads
\begin{align}
\hat U_j^\dag(t) \, \hat{\mathbb{X}} \, \hat U_j(t) = \mathbf{S}_j(t) \hat{\mathbb{X}}. 
\end{align}
This means that the coefficients $F_j$ that we defined in Eq.~\eqref{chap:decoupling:time:evolution:ansatz} have a one-to-one relationship with those in Eq.~\eqref{eq:phase:space:ansatz}.

\subsection{Proof of the phase-space decoupling theorem}
Here we prove the decoupling theorem in phase space. We start by extending the Hamiltonian matrix to include all $m$ elements in the algebra, such that
\begin{align}
\mathbf{H}(t) = \sum_{j = 1}^m G_j(t)  \mathbf{H}_j.
\end{align}
We then differentiate the ansatz in Eq.~\eqref{eq:phase:space:ansatz} with respect to time $t$. We find
\begin{align} \label{chap:decoupling:eq:differentiated:U}
\frac{d \mathbf{S}(t)}{dt} =   \sum_{j = 1}^n \dot{F}_j (t) \left( \prod_{k = 1}^{j - 1} \exp[ \, F_k\, \mathbf{\Omega} \mathbf{H}_k] \right) \, \mathbf{\Omega} \mathbf{H}_j \, \left( \prod_{k = j}^n \exp[ \, F_k \, \mathbf{\Omega} \mathbf{H}_k] \right) .
\end{align}
We then multiply by the inverse operator $\mathbf{S}^{-1}(t)$ on the right-hand-side and use the fact that $d \mathbf{S}(t) /dt = \mathbf{\Omega} \mathbf{H}(t) \,\mathbf{S}(t)$ to find
\begin{align} \label{chap:decoupling:eq:part:solved}
\sum_{j = 1}^n G_j(t) \,\mathbf{\Omega} \mathbf{H}_j &=  \,  \sum_{j = 1}^n \dot{F}_j (t) \left( \prod_{k = 1}^{j - 1} \exp[ \, F_k \, \mathbf{\Omega} \mathbf{H}_k ] \right) \, \mathbf{\Omega} \mathbf{H}_j \, \left( \prod_{k = j - 1}^1 \exp[ -  \, F_k \, \mathbf{\Omega}\mathbf{H}_k ] \right)
\end{align}
By again defining 
\begin{equation}
(ad \boldsymbol{X}) \boldsymbol{X} = [\boldsymbol{X},\boldsymbol{Y}]  ,
\end{equation}
we can use Lemma 1 and Lemma 2 in the main text to arrive at an expression very similar to the one in Eq.~\eqref{chap:decoupling:eq:differential:equation:sum}:
\begin{equation} 
\sum_{k = 1}^n G_k (t) \, \mathbf{\Omega}\mathbf{H}_k =  \, \sum_{j = 1}^n \sum_{k = 1}^n \dot{F}_j (t) \, \chi_{kj} \, \mathbf{\Omega} \mathbf{H}_k  .
\end{equation}
This again shows us that the algebra elements are not the Hamiltonian matrices themselves, but rather the products of the Hamiltonian matrices with the symplectic form.
Then, using the linear independence of the matrices, we find
\begin{equation} 
\sum_{k = 1}^n G_k (t) \,  =  \, \sum_{j = 1}^n \sum_{k = 1}^n \dot{F}_j (t) \, \chi_{kj} .
\end{equation}
For quadratic systems, this method is equivalent to that presented in Section~\ref{sec:decoupling}. Its advantage is that it reduces a complicated operator-based multiplication by congruence to that of simple matrix multiplication. Whether this phase space treatment of the Hilbert space treatment is preferable depends on the problem.

\section{Calculations for Section~4: Hamiltonian with linear terms} \label{app:linear:terms:calculation}
In this appendix, we compute the $F$-coefficients in Eq.~\eqref{eq:linear:F:coefficients}.
First, we note that for arbitrary complex numbers $\theta_+ $ and $\theta_-$, the following relations holds:
\begin{align} \label{eq:BCH:linear}
e^{\theta_+ \hat a^\dag - \theta_-^* \hat a} = e^{\theta_+ \hat a^\dag} e^{- \theta_-^* \hat a} e^{- \frac{1}{2} \theta_+ \theta_-^*} = e^{- \theta_-^* \hat a} e^{\theta_+ \hat a^\dag} e^{\frac{1}{2} \theta_+ \theta_-^*} .
\end{align}
For $\theta_+ = \theta_- $ this expression reduces to the familiar relation for displacement operators $D(z) = e^{z\hat a^\dag - z^*\hat a }=e^{z\hat a^\dag }e^{-z^*\hat a }e^{-|z|^2/2}$. It also directly follows from Eq.~\eqref{eq:BCH:linear} that
\begin{align} \label{app:eq:displacements}
e^{-\theta \hat a^\dag}  \, \hat a \, e^{\theta \hat a^\dag} & = \hat a + \theta, \nonumber \\
e^{\theta \hat a}  \, \hat a^\dag \, e^{-\theta \hat a } & = \hat a^\dag + \theta .
\end{align}
The relations in Eq.~\eqref{app:eq:displacements} are often referred to as displacements of the annihilation and displacement operators, which do not alter the commutator relation $[\hat a,\hat a^\dag] = 1$. In contrast, the number operator $\hat a^\dag \hat a$ induces rotations of the creation and annihilation operators. We also not that any function $f(\hat a,\hat a^\dag)$ that can be Taylor-expanded changes according to
\begin{equation} \label{app:eq:rotation:relation}
e^{-\alpha \, \hat a^\dag \hat a } f(\hat a ,\hat a^\dag) e^{\alpha \, \hat a^\dag  \hat a } =f(\hat a  e^{\alpha},\hat a^\dag  e^{-\alpha}) \, ,
\end{equation}
which holds for arbitrary $\alpha$. These quantities are useful to us going forwards.

\subsection{Deriving the differential equations}
We start by reprinting the ansatz in Eq.~\eqref{eq:linear:ansatz} for the evolution generated by the Hamiltonian in Eq.~\eqref{eq:Hamiltonian:linear:driving}:
\begin{equation} \label{app:eq:linear:ansatz}
\hat U(t) = e^{- i F_0 \hat a^\dag \hat a} \, e^{- i F_+ \hat a^\dag } \, e^{- i F_- \hat a }.
\end{equation}
Our goal now is to determine $F_0$ and $F_\pm$ in Eq.~\eqref{app:eq:linear:ansatz}. We begin by differentiating Eq.~\eqref{app:eq:linear:ansatz} with respect to time $t$ to find
\begin{align} \label{app:eq:diff:linear:ansatz}
\frac{d}{dt} \hat U(t) &= - i  \dot{F}_0 \, \hat a^\dag \hat a \, e^{- i F_0 \hat a^\dag \hat a} \, e^{- i F_+ \hat a^\dag} \, e^{- i F_- \hat a} -i  \dot{F}_+ \, e^{- i F_0 \hat a^\dag \hat a}  \hat a^\dag \, e^{- i F_+ \hat a^\dag} \, e^{- i F_- \hat a}  - i \dot{F}_- \, e^{- i F_0 \hat a^\dag \hat a}\, e^{- i F_+ \hat a^\dag} \, \hat a \, e^{- i F_- \hat a}.
\end{align}
We then multiply Eq.~\eqref{app:eq:diff:linear:ansatz} by $\hat U^{-1}(t)$ on the right to find
\begin{align} \label{app:eq:multiply:Um1}
\frac{d}{dt} \hat U(t) \hat U^{-1}(t)&=- i \dot{F}_0 \, \hat a^\dag \hat a   - i  \dot{F}_+ e^{- i F_0 \hat a^\dag \hat a}  \, \hat a^\dag \,  e^{ i F_0 \hat a^\dag \hat a}  - i   \dot{F}_- e^{- i F_0 \hat a^\dag \hat a} \, e^{- i F_+ \hat a^\dag} \, \hat a \, e^{i F_+ \hat a^\dag} \, e^{i F_0 \hat a^\dag \hat a} .
\end{align}
We then compute the expressions that arise from the congruence multiplications in the second and third term in Eq.~\eqref{app:eq:multiply:Um1}. We find, using the relations in Eq.~\eqref{app:eq:displacements} and Eq.~\eqref{app:eq:rotation:relation},
\begin{subequations}
\begin{align}
e^{- i F_0 \hat a^\dag \hat a} \, \hat a^\dag \, e^{i F_0 \hat a^\dag \hat a} &=   e^{- i F_0 } \hat a^\dag,   \label{eq:linear:congruence:1} \\
e^{- i F_0 \hat a^\dag \hat a} \, \hat a \, e^{i F_0 \hat a^\dag \hat a} &= e^{i F_0 } \hat a , \label{eq:linear:congruence:2} \\
e^{- i F_+ \hat a^\dag } \, \hat a \, e^{i F_+ \hat a^\dag} &= \hat a + i F_+,  \label{eq:linear:congruence:3}\\
e^{- i F_- \hat a } \, \hat a^\dag \, e^{i F_- \hat a}&= \hat a^\dag - i F_- . \label{eq:linear:congruence:4}
\end{align}
\end{subequations}
We do not immediately need Eq.~\eqref{eq:linear:congruence:4}, but it is useful to us later. Using Eqs.~\eqref{eq:linear:congruence:1},~\eqref{eq:linear:congruence:2}, and~\eqref{eq:linear:congruence:3}, we are able to write
\begin{align}
\frac{d}{dt} \hat U(t) \hat U^{-1}(t)&= - i  \dot{F}_0 \, \hat N_0 - i  \dot{F}_+  e^{- i F_0 } \hat a^\dag  - i   \dot{F}_- \left(e^{i F_0 } \, \hat a + i F_+ \right) .
\end{align}
Then, we set this expression equal to $- i \hat H$, where $\hat H$ is the Hamiltonian with linear terms in Eq.~\eqref{eq:Hamiltonian:linear:driving}. We use linear independence of the operators to identify the following differential equations:
\begin{align} \label{app:eq:linear:drive:diff:eqs}
&1 = \dot{F}_0,
&&g_+ (t)  = \dot{F}_+ \, e^{- i F_0 },
&&g_-(t)  = \dot{F}_- \, e^{i F_0 }.
\end{align}
The first equation can be straight-forwardly solved to find $F_0 = t$ with boundary condition $\dot{F}_0(t = 0) = 0$, which corresponds to the free evolution of the harmonic oscillator. We then rearrange the last two equations in Eq.~\eqref{app:eq:linear:drive:diff:eqs} to find the following solutions:
\begin{align} \label{app:eq:linear:F:coefficients}
 &F_+= \int^t_0 \mathrm{d}t^\prime \, g_+(t') \, e^{i t'},  &&
F_- =\int^t_0 \mathrm{d}t^\prime \, g_-(t') e^{- i t'} .
\end{align}

\subsection{Computing the quadratures}
We now derive the expressions for the time-evolution of the quadratures $\hat X(t)$ and $\hat P(t)$, which are shown in Eq.~\eqref{eq:computed:quadratures}. Using the expressions in Eqs.~\eqref{eq:linear:congruence:1}--\eqref{eq:linear:congruence:4}, we find that
\begin{align}
\hat X(t) &= \hat U^\dag(t) \, \hat X \, \hat U(t) \nonumber \\
&= \frac{1}{\sqrt{2}} \left( \hat U^\dag(t)  \, \hat a^\dag \, \hat U(t) + \hat U^\dag(t) \, \hat a  \,\hat U(t) \right) \nonumber \\
&= \frac{1}{\sqrt{2}} \left[  e^{ i F_0 } \left( \hat a^\dag  + i F_-\right) + e^{- i F_0} \left( \hat a - i F_+ \right) \right].
\end{align}
Similarly,
\begin{align}
\hat P(t) &= \hat U^\dag(t) \, \hat P \, \hat U(t) \nonumber \\
&= \frac{i}{\sqrt{2}} \left( \hat U^\dag(t)  \, \hat a^\dag \, \hat U(t) - \hat U^\dag(t) \, \hat a  \,\hat U(t) \right) \nonumber \\
&= \frac{i}{\sqrt{2}} \left[  e^{ i F_0 } \left( \hat a^\dag  + i F_-\right) - e^{- i F_0} \left( \hat a - i F_+ \right) \right].
\end{align}
For an initially coherent state $\ket{\alpha}$, we find that the expectation values are given by
\begin{equation}
\begin{split}
\braket{\hat X(t)} &=  \frac{1}{\sqrt{2}} \left[  e^{ i F_0 } \left( \alpha^*  + i F_-\right) + e^{- i F_0} \left( \alpha - i F_+ \right) \right],  \\
\braket{\hat P(t)} &= \frac{i}{\sqrt{2}} \left[  e^{ i F_0 } \left( \alpha^* + i F_-\right) - e^{- i F_0} \left( \alpha  - i F_+ \right) \right].
\end{split}
\end{equation}

\section{Calculation for Section~5: Hamiltonian with quadratic terms} \label{app:quadratic:terms:calculation}

Here we derive the differential equations shown in Eq.~\eqref{sec:quadratic:drive}.
We start by reprinting the ansatz in Eq.~\eqref{eq:quadratic:ansatz} for the evolution generated by the quadratic Hamiltonian in Eq.~\eqref{eq:quantum:harmonic:oscillator}
\begin{align} \label{eq:app:quadratic:U}
\hat U(t) = e^{- i \xi_+ \hat K_+} \, e^{- i\xi_0 \hat K_0} \, e^{- i \xi_- \hat K_-},
\end{align}
where the operators $\hat K_0, \hat K_\pm$ are given in Eq.~\eqref{eq:K:algebra} and where $\xi_0, \xi_\pm$ are the coefficients we wish to solve for as a function of time.

We start by differentiating Eq.~\eqref{eq:app:quadratic:U} with respect to time $t$ to find:
\begin{align} \label{app:eq:quadratic:first:step}
\dot{\hat{U}}(t) = - i \dot{\xi}_+ \, \hat K_+\, e^{- i\xi_+ \hat K_+}  \, e^{- i \xi_0 \hat K_0} \, e^{- i \xi_- \hat K_-} - i \dot{\xi}_0 \, e^{- i \xi_+ \hat K_+} \,\hat K_0  \, e^{ - i\xi_0 \hat K_0} \, e^{- i \xi_- \hat K_-} - i  \dot{\xi}_- \,e^{- i \xi_+ \hat K_+}  \, e^{- i\xi_0 \hat K_0} \, \hat K_- \, e^{- i \xi_- \hat K_-}.
\end{align}
We then multiply Eq.~\eqref{app:eq:quadratic:first:step} by $\hat U^{-1}(t)$ on the right to find
\begin{align} \label{app:eq:quadratic:midstep}
\dot{\hat{U}}(t) \hat U^{-1}(t)  =  - i  \dot{\xi}_+ \hat K_+  - i  \dot{\xi}_0  \, e^{- i \phi_+ \hat K_+} \hat K_0  \, e^{i \xi_+  \hat K_+}  - i  \dot{\xi}_- e^{- i \xi_+ \hat K_+}  \, e^{-i \xi_0 \hat K_0} \, \hat K_- \, e^{i  \xi_0 \hat K_0} \, e^{i  \xi_+ \hat K_+} .
\end{align}
We then see from the last two terms in Eq.~\eqref{app:eq:quadratic:midstep}, that we need to compute the following multiplications by congruence. We find:
\begin{align}
e^{- i \xi_+ \hat K_+} \hat K_0  \, e^{i \xi_+  \hat K_+} &= \hat K_0 + i  \hat K_+ \xi_+ ,  \nonumber \\
 e^{- i \xi_0 \hat K_0} \, \hat K_- \, e^{i\xi_0 \hat K_0} & = e^{i \xi_0} \hat K_-,  \\
 e^{- i \xi_+ \hat K_+}  \, \hat K_- \, e^{i \xi_+ \hat K_+}   &=  \hat K_- +  2 i \hat K_0 \xi_+ - \hat K_+ \xi_+^2. \nonumber
\end{align}
Thus, after dividing Eq.~\eqref{app:eq:quadratic:midstep} by $-i $ and rearranging, we find
\begin{align}
\dot{\hat{U}}(t) \hat U^{-1}(t) 
&= \left( \dot{\xi}_+ + i \dot{\xi}_0  \xi_+  - \dot{\xi}_- \xi_+^2 e^{i \xi_0} \right) \hat K_+  + \left( \dot{\xi}_0 + 2 i  \dot{\xi}_- \xi_+ e^{i \xi_0} \right) \hat K_0 + \dot{\xi}_- e^{i \xi_0}  \hat K_-. 
\end{align}
We then set this expression equal to $- i \hat H$, where $\hat H$ is the quadratic Hamiltonian in Eq.~\eqref{eq:quantum:harmonic:oscillator}. By using the linear independence of the operators, we are able to drive the following three differential equations:
\begin{align} \label{app:eq:quadratic:diff:eqs}
\lambda_+ &= \dot{\xi}_+ + i  \dot{\xi}_0  \xi_+  - \dot{\xi}_- \xi_+^2 e^{i \xi_0},  \nonumber \\
1 &=  \dot{\xi}_0 + 2 i  \dot{\xi}_- \xi_+ e^{i \xi_0},   \\
\lambda_- &= \dot{\xi}_- e^{i \xi_0}. \nonumber
\end{align}
where we recall that $\lambda_0$ and $\lambda_\pm$ are dimensionless functions of time that appear in the Hamiltonian in Eq.~\eqref{eq:quantum:harmonic:oscillator}.

We can then rearrange the equations in Eq.~\eqref{app:eq:quadratic:diff:eqs} to isolate the derivatives. We start by noting that the third equation implies that $\dot{\xi}_- = \lambda_- e^{- i \xi_0}$.
This allows us to rewrite the first and second equations in Eq.~\eqref{app:eq:quadratic:diff:eqs} as
\begin{align} \label{app:eq:quadratic:diff:eqs:rearranged}
\lambda_+ &= \dot{\xi}_+ + i  \dot{\xi}_0  \xi_+  - \lambda_-  \xi_+^2 ,  \nonumber \\
1 &=  \dot{\xi}_0 + 2 i  \lambda_-  \xi_+ .
\end{align}
We then rearrange the second equation in Eq.~\eqref{app:eq:quadratic:diff:eqs:rearranged} to find $ \dot{\xi}_0 =1  - 2 i \lambda_- \xi_+ $. Inserting this into the first equation in Eq.~\eqref{app:eq:quadratic:diff:eqs:rearranged} and rearranging again, we find
\begin{align}
\dot{\xi}_+ = \lambda_+ - i  \xi_+ - \lambda_- \xi_+^2.
\end{align}
In summary, the differential equations for $\xi_0$, $\xi_+$, and $\xi_-$ are given by
\begin{align} \label{app:eq:quadratic:diff:eqs}
\dot{\xi}_+& = \lambda_+(t) - i  \xi_+ - \lambda_-(t) \xi_+^2,  \nonumber \\
\dot{\xi}_0    &=  2 i  \lambda_-(t) \xi_+ - 1 ,   \\
\dot{\xi}_-&=  \lambda_-(t) \,  e^{ - i \xi_0},  \nonumber
\end{align}
where we have restored the potentially explicit time-dependence of $\lambda_\pm(t)$.

\subsection{Solving the differential equations for constant coefficients}
When the coefficients $\lambda_\pm(t) $ are constant in time with $\lambda_\pm(t) = \lambda_\pm$, we can find an exact solution to the differential equation in Eq.~\eqref{app:eq:quadratic:diff:eqs}.
Using a standard symbolic solver like \textit{Mathematica}, we find the solution
\begin{align} \label{app:eq:first:solution}
\xi_+(t) =\frac{i}{2\lambda_-}   \left\{  2 \Gamma \tan\left[   \tan^{-1}\left(\frac{1}{2\Gamma} \right)   - i t \Gamma \right]	-1 \right\} ,
\end{align}
where we have defined
\begin{align}
\Gamma^2 =  \lambda_+ \lambda_-  - \frac{1}{4} .
\end{align}
To simplify Eq.~\eqref{app:eq:first:solution}, we consider the following addition formula 
\begin{equation}
\tan(A+ B) = \frac{\tan(A) + \tan(B)}{1 - \tan(A) \tan(B)}.
\end{equation}
This, and noting that  $\tan(i A) = i \tanh(A)$, allows us to write Eq.~\eqref{app:eq:first:solution} as
\begin{equation} \label{app:eq:xi:intermediary:2}
\xi_+ = \frac{i}{2\lambda_-}   \left[   \frac{1  - 2 i \Gamma\tanh( t \Gamma) }{1 + \frac{1}{2\Gamma} i \tanh( t \Gamma) } - 1 \right]
\end{equation}
We then multiply out the denominator in Eq.~\eqref{app:eq:xi:intermediary:2} and rearrange the expression to find
\begin{equation} \label{app:eq:xi:intermediary:3}
\xi_+ =
\frac{1}{2\lambda_-}   \left[   \frac{  2  \Gamma\tanh( t \Gamma)  + \frac{1}{2\Gamma} \tanh( t \Gamma)}{1 + \frac{1}{2\Gamma} i \tanh( t \Gamma) } \right].
\end{equation}
Finally, we note that $ \frac{1}{2} + 2  \Gamma^2 =  \frac{1}{2} + 2  \left(  \lambda_+ \lambda_- -  \frac{1}{4} \right) = 2 \lambda_+ \lambda_- $, which means that we can write Eq.~\eqref{app:eq:xi:intermediary:3} as 
\begin{equation} \label{app:eq:xi+:result}
\xi_+ = \frac{\lambda_+}{ \Gamma}   \left[   \frac{  \sinh( t \Gamma)}{\cosh(t \Gamma) + \frac{i}{2\Gamma}  \sinh( t \Gamma) } \right]  .
\end{equation}
which is the final result. The solution for $\xi_0$ can be tested by inserting the result in Eq.~\eqref{eq:quadratic:algebra:constant:solutions} into the second equation in Eq.~\eqref{app:eq:quadratic:diff:eqs}. The result satisfies the differential equation. The same can be done for $\xi_-$.

\section{Calculations for Section~6: Most general Gaussian Hamiltonian} \label{app:linear:quadratic:terms}
In this appendix, we provide the details for the calculations in Section~\ref{sec:linear:quadratic:drive}.
Our main goal is to compute the expression in Eq.~\eqref{eq:evolving:Hamiltonian}. To do so, we need to determine how the linear term $g_+\hat a^\dag + g_- \hat a$ evolves in the frame rotating with the quadratic algebra defined in Eq.~\eqref{eq:K:algebra}. More specifically, we wish to derive an expression for the term
\begin{align} \label{app:eq:term:to:transform}
\hat U_Q^{-1} (t) \, \left[ g_+(t)\hat a^\dag + g_- (t) \hat a \right] \hat U_Q(t).
\end{align}
To do so, we first need to compute the following expressions:
\begin{equation}  \label{app:eq:congruence:relations}
\begin{split} 
e^{i \xi_0 \hat K_0 } \, \hat a \, e^{- i \xi_0 \hat K_0} &= e^{- \frac{1}{2}i \xi_0 } \, \hat a , \\
e^{i \xi_0 \hat K_0 } \, \hat a^\dag \, e^{- i \xi_0 \hat K_0} &= e^{\frac{1}{2}i \xi_0} \, \hat a^\dag,  \\
e^{i \xi_+ \hat K_+ } \, \hat a \, e^{- i \xi_+ \hat K_+} &= \hat a - i \xi_+ \hat a^\dag,  \\
e^{i \xi_- \hat K_- } \, \hat a^\dag \, e^{- i \xi_- \hat K_-} &=  \hat a^\dag + i \xi_- \hat a .
\end{split}
\end{equation}
Quadratic transformations of the kind shown in the second two equation of Eq.~\eqref{app:eq:congruence:relations} are also known as Bogoliubov transformations. They map ladder operators to linear mixtures of themselves. Using the expressions in Eq.~\eqref{app:eq:congruence:relations}, we find 
\begin{equation}
\begin{split}
e^{i \xi_- \hat K_-} \, e^{i \xi_0 \, \hat K_0} \, e^{i \xi_+ \, \hat K_+} \, \hat a \, e^{- i \xi_+ \hat K_+} \, e^{- i \xi_0 \, \hat K_0} \, e^{- i \xi_- \hat K_-} &= \left( e^{- \frac{1}{2} i \xi_0} + \xi_+ \xi_- \, e^{\frac{1}{2}i \xi_0}  \right)  \hat a - i \xi_+ \, e^{\frac{1}{2}i  \xi_0}  \hat a^\dag    \\
e^{i \xi_- \hat K_-} \, e^{i \xi_0 \, \hat K_0} \, e^{i \xi_+ \, \hat K_+} \, \hat a^\dag \, e^{- i \xi_+ \hat K_+} \, e^{- i \xi_0 \, \hat K_0} \, e^{- i \xi_- \hat K_-} &= e^{\frac{1}{2} i \xi_0} \left( \hat a^\dag  + i \xi_- \hat a \right).
\end{split}
\end{equation}
The term in Eq.~\eqref{app:eq:term:to:transform} therefore becomes
\begin{align}
\hat U_Q^{-1} (t) \, &\left( g_+(t)\hat a^\dag + g_- (t) \hat a \right) \hat U_Q(t) = g_+(t) e^{\frac{1}{2} i \xi_0} \left( \hat a^\dag  + i \xi_- \hat a \right) + g_- (t) \left[ \left( e^{- \frac{1}{2} i \xi_0} + \xi_+ \xi_- \, e^{\frac{1}{2} i\xi_0}  \right)  \hat a - i \xi_+ \, e^{\frac{1}{2}i\xi_0}  \hat a^\dag   \right],   \nonumber \\
&= \left[ g_+(t) e^{\frac{1}{2} i \xi_0}  - i g_-(t) \xi_+ e^{\frac{1}{2} i \xi_0}\right] \hat a^\dag+ \left[  i g_+(t) \xi_-  e^{\frac{1}{2} i \xi_0}  + g_-(t) \left( e^{- \frac{1}{2} i \xi_0} + \xi_+ \xi_- e^{\frac{1}{2} i \xi_0 } \right) \right] \hat a .
\end{align}
For convenience, we then define the following coefficients
\begin{equation}
\begin{split}
\mu(t) &= g_+(t) e^{\frac{1}{2} i \xi_0}  - i g_-(t) \xi_+ e^{\frac{1}{2} i \xi_0},   \\
\nu (t) &=  i g_+(t) \xi_-  e^{\frac{1}{2} i \xi_0}  + g_-(t) \left( e^{- \frac{1}{2} i \xi_0} + \xi_+ \xi_- e^{\frac{1}{2} i \xi_0 } \right) .
\end{split}
\end{equation}
which allows us to write
\begin{align}
\hat U_Q^{-1} (t) \, &\left[ g_+(t)\hat a^\dag + g_- (t) \hat a \right] \hat U_Q(t) = \mu (t) \hat a^\dag + \nu(t) \hat a.
\end{align}
Following the same procedure as outlined in~\ref{app:linear:terms:calculation}, we now make the ansatz for the linear algebra:
\begin{align}
\hat U_L(t) = e^{- i F_+ \hat  a^\dag } \, e^{- i F_- \hat a },
\end{align}
where the two coefficients $F_\pm$ are given by
\begin{align} \label{app:eq:linear:F:coefficients}
 &F_+= \int^t_0 \mathrm{d}t^\prime \, \nu(t') \, e^{i t'},  &&
F_- =\int^t_0 \mathrm{d}t^\prime \, \mu(t') e^{- i t'} .
\end{align}
These integrals are undoubtedly challenging to compute, but doing so allows us to solve the dynamics of the most general Gaussain single-mode Hamiltonian.

\footnotesize
\bibliographystyle{iopart-num}
\bibliography{thesis_bibliography}

\end{document}